\documentclass[desactivate]{aa}

\usepackage{graphicx}
\usepackage{txfonts}
\usepackage{color}

\begin{document}

   \title{SPT: Spectral Transformer for Red Giant Stars Age and Mass Estimation}

   \author{
      Mengmeng Zhang\inst{1}\
      \and
      Fan Wu\inst{1}\thanks{These authors contributed equally to this work.} 
      \and
      Yude Bu\inst{1}\thanks{Corresponding author.}
      \and
      Shanshan Li\inst{1}
      \and
      Zhenping Yi\inst{2}
      \and
      Meng Liu\inst{2}
      \and
      Xiaoming Kong\inst{2}
  }
  
  \institute{
      School of Mathematics and Statistics, Shandong University, Weihai, 264209, Shandong, People's Republic of China\\
      \email{buyude@sdu.edu.cn}
      \and
      School of Mechanical, Electrical and Information Engineering, Shandong University, Weihai, 264209, Shandong, People's Republic of China\\
  }  

   \date{Received September 15, 1996; accepted March 16, 1997}

 
  \abstract
  {The age and mass of red giants are essential for understanding the structure and evolution of the Milky Way. 
   Traditional isochrone methods for these estimations are inherently limited due to overlapping isochrones in the Hertzsprung-Russell 
   diagram, while asteroseismology, though more precise, requires high-precision, long-term observations. In response to these 
   challenges, we developed a novel framework, Spectral Transformer (SPT), to predict the age and mass of red giants aligned with 
   asteroseismology from their spectra.  A key component of SPT, the Multi-head Hadamard Self-Attention mechanism, 
   designed specifically for spectra, can capture complex relationships across different wavelength. Further, we introduced 
   a Mahalanobis distance-based loss function to address scale imbalance and interaction mode loss, and incorporated Monte Carlo 
   dropout for quantitative analysis of prediction uncertainty. Trained and tested on 3,880 red giant spectra from LAMOST, the SPT 
   achieved remarkable age and mass estimations with average percentage errors of 17.64\% and 6.61\%, respectively, and provided 
   uncertainties for each corresponding prediction. The results significantly outperform those of traditional machine learning 
   algorithms and demonstrate a high level of consistency with asteroseismology methods and isochrone fitting techniques. 
   In the future, our work will leverage datasets from the Chinese Space Station Telescope and the Large Synoptic Survey 
   Telescope to enhance the precision of the model and broaden its applicability in the field of astronomy and astrophysics.}


   \maketitle
%

\section{Introduction}\label{sec:intro}
   The precise determination of age and mass for a large sample of stars is essential for a comprehensive understanding of stellar 
   populations and the assembly history of the Galaxy. 
   Red giants, given their unique theoretical properties, are potent instruments for analyzing galactic structures. 
   Their high luminosity, coupled with broad spectrum in age, mass, chemical composition, and evolutionary state serve to enhance their 
   detectability across significant distances \citep{10.1111/j.1365-2966.2005.08689.x, miglio2012red, vrard2022evidence}. 
   This detectability persists even in regions with high extinction, such as the Galactic bulge.
   Thus, red giants emerge as indispensable tools for studying the structure and constituents of the Milky Way.

   The isochrone fitting technique, widely employed to infer the age and mass of stars, compares a star's position on the 
   Hertzsprung-Russell (H-R) diagram against theoretical isochrones.
   These isochrones are derived from stellar evolution models 
   \citep{2014A&A...565A..89B, sanders2018isochrone, squicciarini2022madys,higgins2023stellar}.
   However, the overlap of these isochrones on the H-R diagram complicates the determination of the optimal isochrone for a specific red giant. 
   In addition, both stellar evolution models and isochrones contain certain assumptions and uncertainties that can affect the estimation of 
   age and mass.
   For instance, stellar evolution models often operate under the assumption that stars are isolated, without mass loss or exchange. 
   These processes, however, can and do occur \citep{doi:10.1146/annurev-astro-081309-130806, montalban2021chronologically}. 
   Furthermore, the models often assume knowledge of the initial composition of the star, which is typically unknown 
   \citep{pietrinferni2021updated}. 
   In summary, considering these limitations, relying solely on the isochrone method for the age and mass of red giants estimations seems 
   imprudent.

   By contrast, asteroseismology offers a powerful tool to probe the internal structure and evolution of stars by analysing their oscillation 
   modes. 
   Unlike other methods that rely on external properties like luminosity or color index, asteroseismic ages are derived from the 
   physical state inside a star, thereby reducing certain systematic errors and uncertainties. 
   Using asteroseismology to infer the age and mass of red giants has been proven an effective method 
   \citep{silva2018confirming,wu2018mass, pinsonneault2018second, sun2020mapping, miglio2021age,
   wu2023timing}. 
   However, this method also faces some limitations, such as the need for high-precision and long-term photometric observations, which are 
   difficult to obtain for distant or faint red giants. 
   Moreover, a deep understanding of red giant oscillation theory is essential for the accurate interpretation of observed frequencies and 
   amplitudes.

   Recently, significant advances in the estimation of red giant star ages and masses have been achieved by integrating asteroseismic 
   parameters from the $Kepler$ mission \citep{borucki2010kepler} with spectroscopic parameters from both the APOGEE 
   \citep{2010IAUS..265..480M, 2017AJ....154...94M,jonsson2020apogee} and LAMOST surveys \citep{2012RAA....12..735D,2012RAA....12..723Z,
   Luo_2015,zhang2020deriving}. 
   Previous efforts included various methods. 
   \cite{2016MNRAS.456.3655M} employed a linear regression model to estimate the ages and masses of 1,475 red giants from the APOGEE-Kepler 
   Asteroseismology Science Consortium (APOKASC) catalog \citep{2014ApJS..215...19P}, achieving relative errors of 40\% and 14\%, 
   respectively; \cite{2016ApJ...823..114N} inferred the parameters for 1,639 red giants by fitting polynomials to each pixel of the stellar 
   spectra with APOGEE, achieving mass precision of about 0.07 dex and age precision of roughly 0.2 dex (40\%); \cite{2018MNRAS.475.3633W} 
   trained a machine learning method based on kernel principal component analysis to analyze LAMOST spectral data, and estimated the ages and 
   masses of red giant branch (RGB) stars with typical uncertainties of a few percent for mass and approximately 24\% for age. 
   \cite{2019MNRAS.484..294D} applied a Bayesian artificial neural network to generate red giant estimates from combined astrometric, 
   photometric, and spectroscopic data, with fractional uncertainties less than 10\% for mass and between 10\% and 25\% for age.
   \cite{10.1093/mnras/stad1272} demonstrated the application of a variational autoencoder-decoder in APOGEE and \textit{Kepler} observations. Their 
   method aims to reduce the dimensionality of APOGEE spectra and eliminate abundance information, improving the estimation of spectral ages 
   (by approximately 22\% overall).

   Despite these successes, current methods are greatly influenced by uncertainties and heavily rely on precise spectroscopic parameters 
   including surface gravity $\log g$, effective temperature $T_{\text{eff}}$, metallicity $\text{[Fe/H]}$, and the abundance of carbon and 
   nitrogen. These dependencies might lead to the propagation of systematic error. 
   Beyond including the aforementioned parameters, the spectral data inherently contains a wealth of additional information. 
   This richness mitigates the potential for inductive biases that might arise from the explicit extraction of these spectral parameters.
   Therefore, we seek a novel method to directly learn the relationship between stellar spectra and age or mass without any prior knowledge of 
   fundamental physics or stellar parameters. 
   This innovative approach not only seeks to enhance the precision of red giant age and mass estimations, 
   but also offers observational constraints on theoretical models of stellar evolution.

   Deep learning techniques have transformed various fields due to their robust representation learning capabilities.
   The Transformer model, introduced by \cite{2017arXiv170603762V}, stands out with notable success, especially in natural language processing 
   (NLP) and computer vision (CV).
   These successes are largely attributed to its self-attention mechanism, which enables the model to capture dependencies between different parts 
   of a sequence flexibly, irrespective of their positions.
   This ability makes it a powerful tool for capturing complex semantic relationships. 
   Consequently, there have been advancements in NLP tasks like machine translation \citep{2017arXiv170603762V}, question-answering systems 
   \citep{2018arXiv181004805D}, text generation \citep{radford2018improving,radford2019language,brown2020language}, and others, like GPT-4 
   \citep{2023arXiv230308774O}. 
   Additionally, the applications of the Transformer have extended to CV, with models like the Vision Transformer for image classification 
   \citep{2020arXiv201011929D}, and more exploration in object detection and image segmentation \citep{reddy2021dall,khan2022transformers}.

   While the successes of deep learning and Transformer models have provided potent tools for handling complex and structured data, efforts to 
   fully integrate these advanced techniques with the intricate characteristics and dependencies of spectra remain relatively limited. 
   To address this challenge, we introduced a novel Spectral Transformer model, SPT, designed to extract features from the spectra of red giants more effectively.
   Specifically, we implement our SPT model on a dataset of 3,880 red giant spectra from LAMOST DR9. 
   The age and mass labels for this dataset come from APOKASC-2 \citep{pinsonneault2018second}. 
   Our SPT achieves state-of-the-art performance, with relative errors of 17.64\% in age prediction and 6.61\% in mass prediction. 

   The organization of this paper is as follows. 
   Section \ref{sec:methods} describes the components of the SPT. 
   Section \ref{sec:data} introduces the selection and processing of the data samples. 
   The results for age and mass estimation are presented in Sect. \ref{sec:result}. 
   Section \ref{sec:valuation} discusses comparative experiments approached from different perspectives. 
   Finally, the conclusions are presented in Sect. \ref{sec:conclusion}.

\begin{figure*} 
   \begin{center}
      \includegraphics[width=1.02\linewidth]{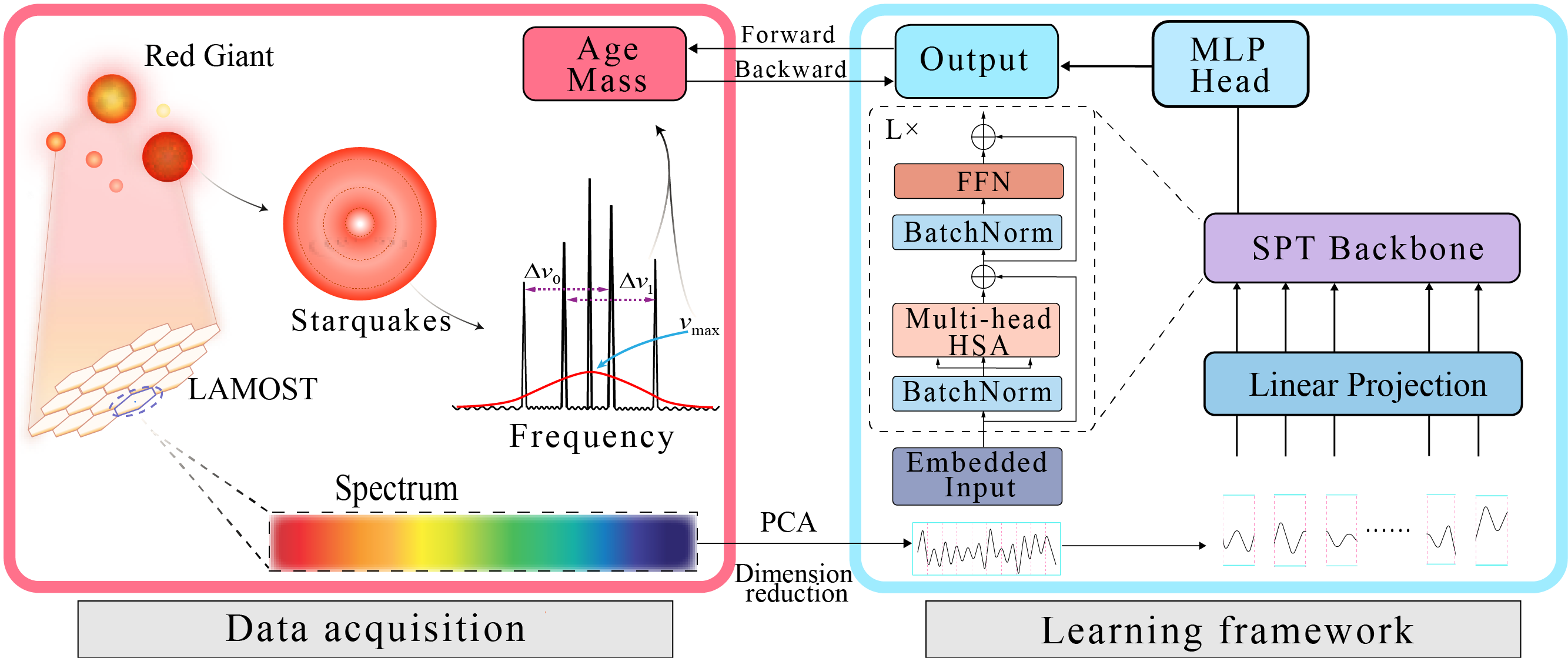}  
   \end{center}
   
  \caption{  Entire framework overview. 
  The data acquisition section (\textit{left panel}) describes the sources of data for the model. 
  This primarily comprises red giant spectra collected by the LAMOST telescope, and the corresponding ages and masses obtained through asteroseismology methods 
  (primarily determined by the mean large frequency, $\Delta\nu $, and the frequency of maximum power, $\nu_{\text{max}}$). 
  These data serve as the foundation for training and testing within the learning framework (\textit{right panel}). 
  The spectra are input into the model after dimensionality reduction via PCA method, while the ages and masses serve as the labels for the model. 
  The input undergoes a transformation through a linear projection layer to generate the embedded input. 
  Subsequently, the SPT backbone is responsible for feature extraction, comprising $L$ SPT blocks (outlined by dashed lines). 
  Each SPT block consists of two BatchNorm (Batch Normalization) layers , a Multi-head HSA layer, a FFN (feedforward network, with two linear layers separated 
  by a GeLU activation), and  $\bigoplus$ (residual connection). 
  The high-level semantic features extracted from the SPT backbone are then fed into the MLP head layer, a fully connected multi-layer neural network. 
  The output is the final result of the model, generated by the MLP head. During the forward propagation process, the model computes the predicted values and loss, 
  while during the backward propagation process, it calculates gradients and updates parameters to optimize predictive performance. }
  \label{fig: modelfigure}
\end{figure*} 

\section{Methods}\label{sec:methods}
   \subsection{Spectral Transformer (SPT)} \label{SPT}
   Our primary goal is to develop an algorithm that captures the intrinsic features of spectra to estimate the age and mass of red giants. 
   The complex and high-dimensional nature of spectra presents a challenge for deep learning methods like Convolutional Neural Networks (CNNs) and Recurrent Neural Networks (RNNs). 
   Traditional CNNs primarily utilize local convolutional kernels for feature extraction, thereby emphasizing local spectral patterns. 
   However, key spectral information might be global or span multiple bands, which could limit CNNs in capturing long-distance 
   dependencies. 
   Recurrent Neural Networks, though adept at processing sequential data, face the problems of vanishing and exploding gradients, especially 
   with long sequences. 
   Moreover, the iterative computation intrinsic to RNNs tend to be less computationally efficient compared to alternative architectures.

   In contrast to CNNs and RNNs, the self-attention mechanism is capable of directly attending to any part of the input sequence, thereby 
   effortlessly capturing long-distance dependencies. 
   For spectra, this implies that the self-attention mechanism can identify correlations and patterns across the entire spectral range without distance constraints. 
   This capability makes the self-attention mechanism particularly suitable for processing high-dimensional and continuous spectra.

   To fully exploit the advantages of the self-attention mechanism, we have designed the Multi-head Hadamard Self-Attention (Multi-head HSA) 
   operator, aiming to process red giant spectra more effectively. 
   The Multi-head HSA, being the core component of our model, combines the global attention power of self-attention and the multi-perspective 
   observation capability of multi-head attention, enabling a more comprehensive capture of information in spectra. 
   A detailed description of the Multi-head HSA is provided in Sect. \ref{Multi-head HSA}.

   Our SPT model follows the workflow illustrated in Fig. \ref{fig: modelfigure}. 
   Initially, red giant spectra collected from the LAMOST telescope and the age and mass obtained through asteroseismic methods serve as the 
   basis for training our model. 
   To address the computational and model training challenges posed by the high-dimensionality of red giant spectra, we employ principal component analysis (PCA) 
   to transform the original spectra from its high-dimensional space to a subspace of 143 dimensions, denoted as n.
   This step not only substantially reduces computational complexity but also ensures that the principal features of the spectral data are 
   retained during the dimensionality reduction process. 
   The dimensionally-reduced spectra are then fed into a linear projection layer to generate the embedded input for the SPT backbone. 
   These backbone comprises multiple SPT blocks, each consisting of a BatchNorm layer, a Multi-head HSA layer, 
   another BatchNorm layer, and a multilayer perceptron (MLP) layer.
   Importantly, the SPT backbone introduces residual connections. 
   The output of the SPT backbone is directly processed by a MLP for processing, which is then converted into the final output for the age 
   and mass of the red giants. 
   It should be noted that the output dimensions of all sub-blocks and the embedding layer within the model are set at $d_\text{model}$.
   Based on experimental results, we configured the SPT backbone with $L=6$, signifying a stack of six SPT blocks.

   \subsection{Multi-head Hadamard Self-Attention}\label{Multi-head HSA}
   To improve the representational capacity of our model for red giant spectra, we introduce a novel attention mechanism: the Multi-head HSA.

\begin{figure*}
      \centering
      \includegraphics[width=0.8\textwidth]{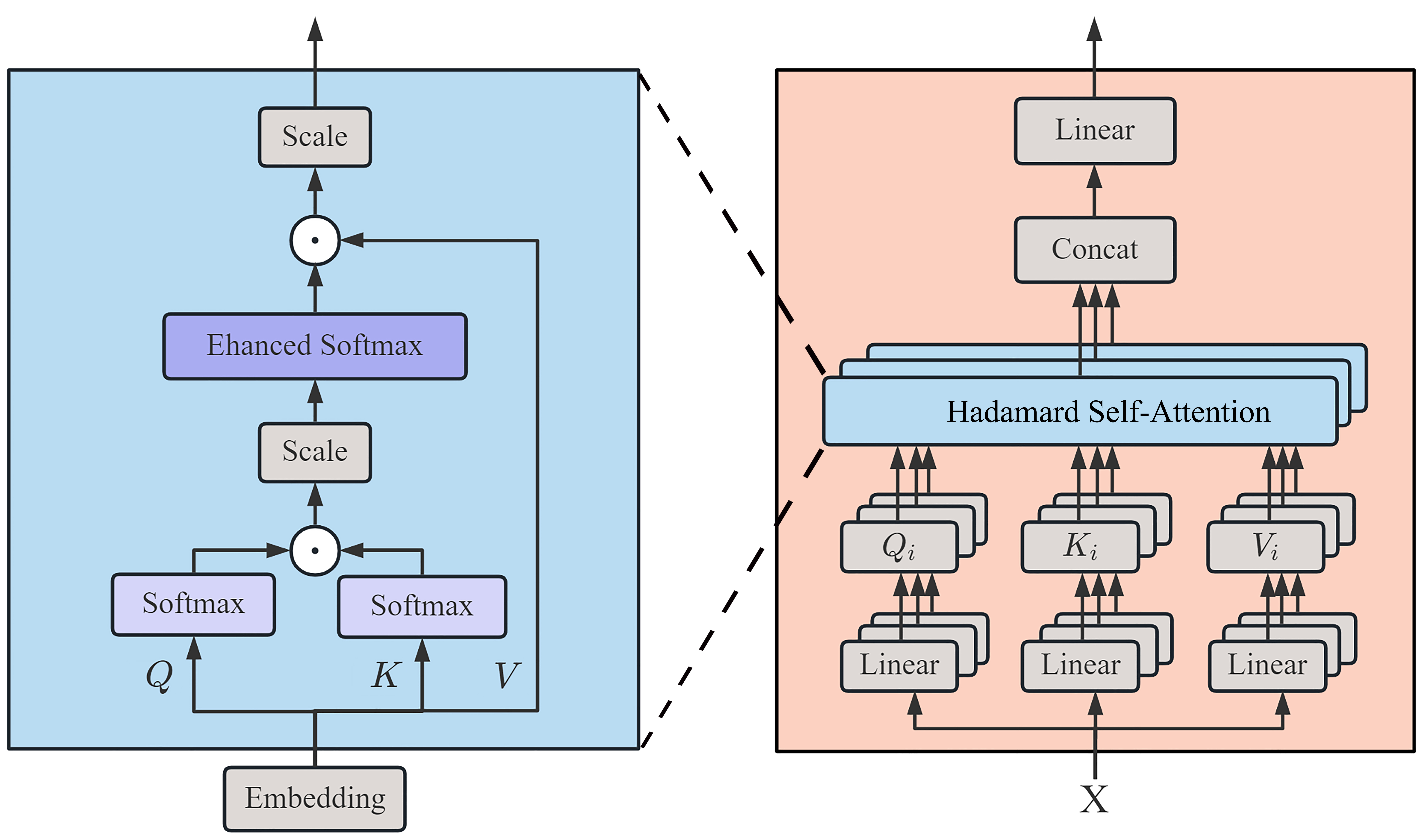}
      
      \caption{Multi-head HSA mechanism. 
      \textit{Left panel:} HSA mechanism, corresponding to the blue box in the \textit{right panel} (Multi-head HSA). 
      In this mechanism, $Q$, $K$, and $V$, representing query, key, and value respectively, are obtained by applying different linear projections to the input, 
      and their dimensions are all $d^{*}=64$ in this paper. 
      \textit{Softmax} is a mathematical function that converts a vector of numbers into a vector of probabilities. 
      $\bigodot$  represents the Hadamard Product. 
      Scale is used to scale the calculated attention scores. 
      \textit{Enhanced softmax} is an improved version of the \textit{softmax} function proposed by us. 
      In the \textit{right panel}, the Linear layer represents a linear projection, X represents the input, and different $Q_i$, $K_i$, $V_i$ are obtained through the 
      linear layer, where $i$ indicates the index of the "head". 
      The "Concat" operation concatenates features from different heads, and then the final result is output through the linear layer. }
      \label{fig:self-attention}
   \end{figure*}

   Firstly, to better understand the Multi-head HSA, we describe our improvements to the self-attention mechanism, as shown in the left panel 
   of Fig. \ref{fig:self-attention}. 
   The fundamental idea of self-attention is to determine the importance distribution of the input by computing its internal relations, i.e., 
   capturing its internal structure by assigning variable weights to each element in the feature vectors extracted from the input. 
   Considering the continuous and high-dimensional nature of red giant spectra, we propose a specific self-attention strategy. 
   In our design, the inputs consist of queries $Q$, keys $K$, and values $V$, 
   all of which are derived through linear projection of the  embedding which is obtained from the previous layer, represented by $X\in\mathbb{R}^{n \times d_\text{model}}$. 
   These components share a common dimension of $d^{*}$, with corresponding parameters $W_i^Q\in\mathbb{R}^{d_\text{model}\times d^*}$, 
   $W_i^K\in\mathbb{R}^{d_\text{model}\times d^*} $, and $W_i^V\in\mathbb{R}^{d_\text{model}\times d^*} $. 
   The output is derived by weighting the values in $V$, with weights determined by the similar attention patterns between $Q$ and $K$. 
   The specific computational process begins by normalizing $Q$ and $K$ using the \textit{softmax} function, 
   followed by calculating their Hadamard product,
   and then scaling it by $d^{*}$. This is followed by an attention enhancement through the \textit{enhanced\_softmax}, and the computation of the 
   Hadamard product with the value $V$, with a final scaling adjustment to complete the process, the computation is provided as follows:
   \begin{align}
   &\text{Attention}(Q,K,V) = \nonumber \\
   &\begin{aligned}
      \textit{enhanced\_softmax} \left[\text{softmax}(Q) \circ \text{softmax}(K)d^*\right] \circ V d^*
   \end{aligned}
   \label{eq1}
   \end{align} 

   Our enhanced self-attention mechanism is characterized by:

   \textbf{Incorporation of Hadamard product:}\quad Traditional attention mechanisms employ dot products to compute the similarity between 
   $Q$ and $K$. 
   However, in our model, we first apply the \textit{softmax} function to $Q$ and $K$, and then use their Hadamard product as a replacement for the dot 
   product. 
   This approach not only reduces the original computational complexity to linear time, 
   but, in contrast to traditional patch-based methods, 
   it offers a more comprehensive consideration of the feature relationships between individual positions within the sequence data.

   \textbf{Scale adjustment:}\quad When calculating weights, we choose to multiply by $d^{*}$ rather than the 
   traditional method of dividing by $\sqrt{d^*}$. 
   As the dimensions of both $Q$ and $K$ are $d^{*}$, 
   this approach adjusts the scale of attention weights back to its original level, enhancing model stability.
   
   \textbf{Attention Augmentation:} \quad We designed an \textit{enhanced\_softmax} function which can further enhance 
   the attention toward continuous spectra. 
   In traditional methods using the \textit{softmax} function, the processed query and key values typically lie between 0 and 1, especially with regularization tricks.
   Even though the exponential computation within the \textit{softmax} function exhibits a "winner-takes-all" characteristic, prioritizing higher values 
   and minimizing lower ones, this prioritization becomes less distinct within the 0-1 range.
   Consequently, there is a risk of an overly uniform attention distribution, preventing the model from emphasizing crucial features effectively. 
   To address these challenges and delineate spectral characteristics more precisely, we introduced an \textit{enhanced\_softmax} function, 
   defined as:
   \begin{align}
      \textit{enhanced\_softmax} = \frac{e^{x_i}-1}{\sum_{j=1}^{n} (e^{x_j}-1)}
      \label{eq0}
   \end{align} 

   More comprehensive discussion and theoretical proof can be found in Appendix \ref{appen A}. 
   The \textit{enhanced\_softmax} function consistently demonstrates superior attention efficiency over the traditional \textit{softmax} function across the 
   entire range $(0, \infty)$. 
   Importantly, it remains unaffected by attention dispersion in the [0,1] interval. 
   This refined \textit{enhanced\_softmax} function better reveals the competitive interplay among features, thus significantly improving the 
   attention distribution of our model when processing spectral data.

   Even though we have improved the self-attention mechanism, a single-head attention mechanism might not fully capture all the patterns and 
   complexities presented in the data. 
   To further enhance model performance, we introduce the Multi-head HSA. 
   The advantage of this method is that it can perform the HSA function in multiple representational subspaces in parallel, 
   enabling the model to simultaneously capturing and focus on information in different representational spaces. 

   \begin{figure*}
      \centering
      \includegraphics[width=\linewidth]{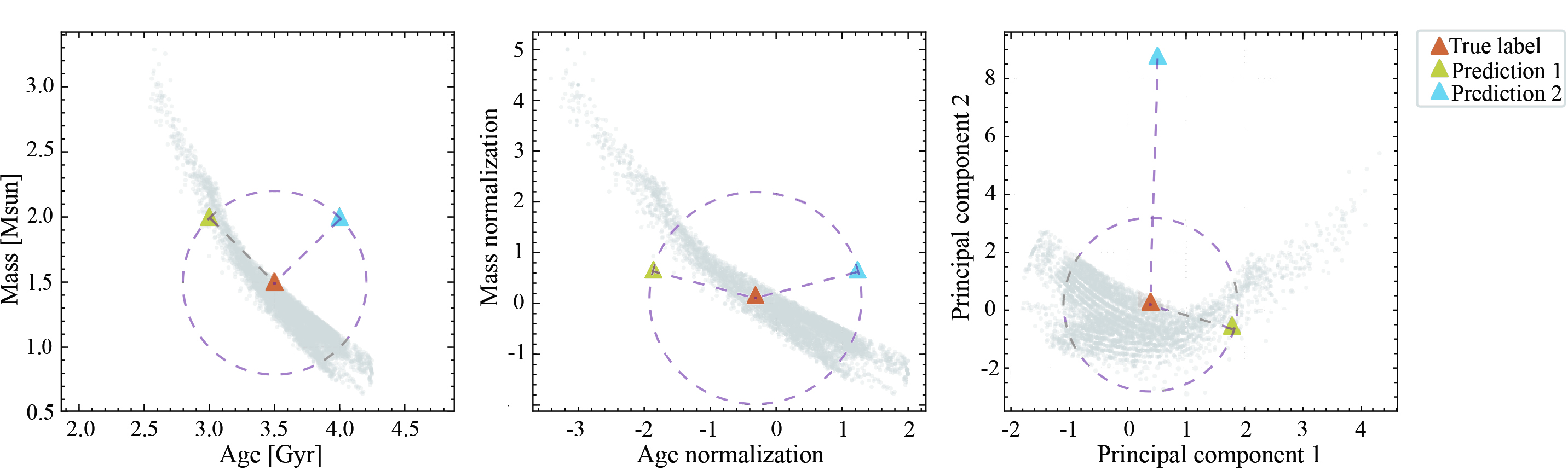}  
      
      \caption{Preference for loss function. 
      To illustrate the effectiveness of different loss functions, we randomly selected a true label ("True label") from the dataset and 
      produced two predictions ("Prediction 1" and "Prediction 2"), represented by red, green, and blue triangles, respectively. 
      The gray dots represent the distribution of data samples. 
      The dashed circle is drawn with a radius equal to the Euclidean distance between the "True Label" and "Prediction 1".
      \textit{Left panel:} Original data distribution. 
      \textit{Middle panel:} Data distribution after z-score normalization is applied separately to age and mass. 
      \textit{Right panel:} Data distribution post-normalization using the Mahalanobis distance.}
      \label{fig:LOSSPRE}
   \end{figure*}
 
   As shown in the right panel of Fig. \ref{fig:self-attention}, the multi-head method employs various self-attention operators, specifically
   the HSA function, as individual "heads" to parallelly extract different features of the spectral data.  
   The operation of each "head" can be represented by Eq. (\ref{eq1}), and the formulaic representation for the complete multi-head processing 
   is given by:
   \begin{align}
   \text{Multi-Head}(Q, K, V)=\text{Concat}(\text{head}_1,...,\text{head}_\text{h})W^O
   \label{eq2}
   \end{align}
   where $W^O\in\mathbb{R}^{hd^*\times d_\text{model}} $ represents the output parameter matrices. 
   The output values of all heads are concatenated and then subjected to a linear projection to obtain the final output.
   In our implementation, we use $h=8$ parallel attention heads, with the dimensions of each head reduced to $d^*=d_\text{model}/h=64$.

   In summary, our proposed Multi-head HSA operator, which combines the Hadamard product and multi-head attention, can more effectively process 
   the information in red giant spectra, thereby enhancing the representational power of our model.

   \subsection{Mahalanobis Distance Loss Function} \label{function}
      In this work, we introduce an integrated framework that directly estimates both the age and mass of red giants from 
      their spectra, rather than constructing two separate models for training and inference. 
      This unified approach not only improves the efficiency of both training and inference but also enhances the robustness and 
      generalization capability of the model by extracting parameter-free features specific to red giants.

      However, the loss function of the model, which combines the losses from both age and mass, inevitably introduces some bottlenecks.
      The first bottleneck arises from the different scales of age and mass data datasets, which might lead to numerical dependencies, 
      compromising the balanced performance of the model across these parameters. 
      To address this, we have considered normalization techniques.

      Traditional normalization methods for age and mass parameters effectively address scale differences, but they introduce a new bottleneck: 
      the loss of interaction patterns, such as relative trends between data pairs. 
      This issue is clearly shown in Fig. \ref{fig:LOSSPRE}. 
      The left panel of Fig. \ref{fig:LOSSPRE} presents the original data distribution.
      Here both "Prediction 1" (green triangle) and "Prediction 2" (blue triangle) have the same Euclidean distances from the "True label" 
      (red triangle).
      However, this distance does not truly represent the underlying data trends and relationships. 
      For example, while "Prediction 1" is within the distribution range of the "True label", "Prediction 2" is an outlier. 
      This suggests that the model generating "Prediction 2" does not capture the main data trend and misses the basic relationship between
      mass and age.
      Therefore, it should be penalized higher. 
      However, when normalizing the parameters separately, the middle panel of Fig. \ref{fig:LOSSPRE} shows no added penalty for "Prediction 2",
      making its distance to the "True label" similar to that of "Prediction 1". 
      This indicates that typical normalization methods do not apply the necessary penalties.

      To overcome these bottlenecks, we introduce a loss function using the Mahalanobis distance. 
      Unlike standard normalization methods, the Mahalanobis distance autonomously adjusts the principal components, tackling the problem of 
      scale differences. 
      Crucially, it maintains data relationships, preventing the loss of key interaction patterns.  
      This benefit is evident in the right panel of Fig. \ref{fig:LOSSPRE}, where the Mahalanobis distance effectively detects and penalizes 
      outliers, represented by the blue data points.

      In particular, our loss function, denoted as $J\left(y_i,\hat{y}_i\right) $, is defined as:
      \begin{align}
      J\left(y_i,\hat{y}_i\right)=\left(y_i-\hat{y}_i\right)^T \Sigma^{-1}\left(y_i-\hat{y}_i\right)
      \label{eq4}
      \end{align}
      where $y_i$ represents the true value in the training set and $\hat{y}_i$ is the prediction from the SPT model. 
      The term $\Sigma$ is the covariance matrix. 
      By incorporating this covariance, our loss function considers the relationships between multi-objective labels 
      in a transformed space, allowing a more meticulous training process.

      \begin{figure*}
         \centering
         \includegraphics[width=0.46\textwidth]{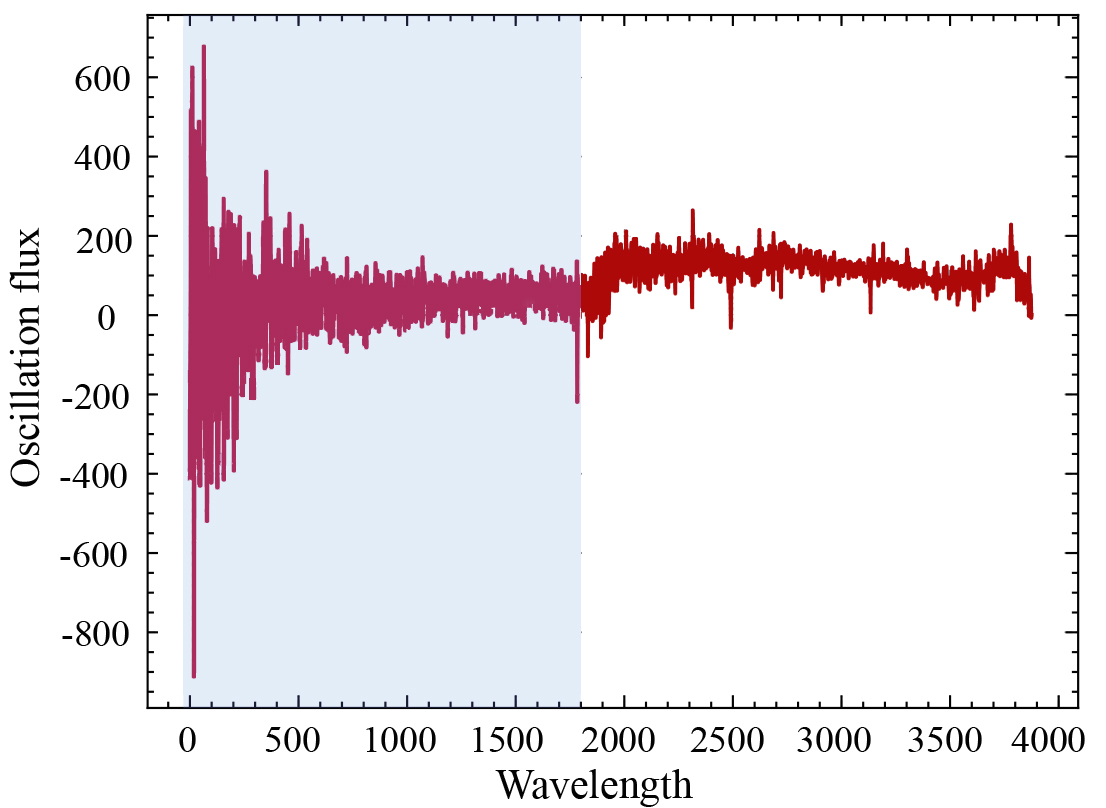}
         \includegraphics[width=0.46\textwidth]{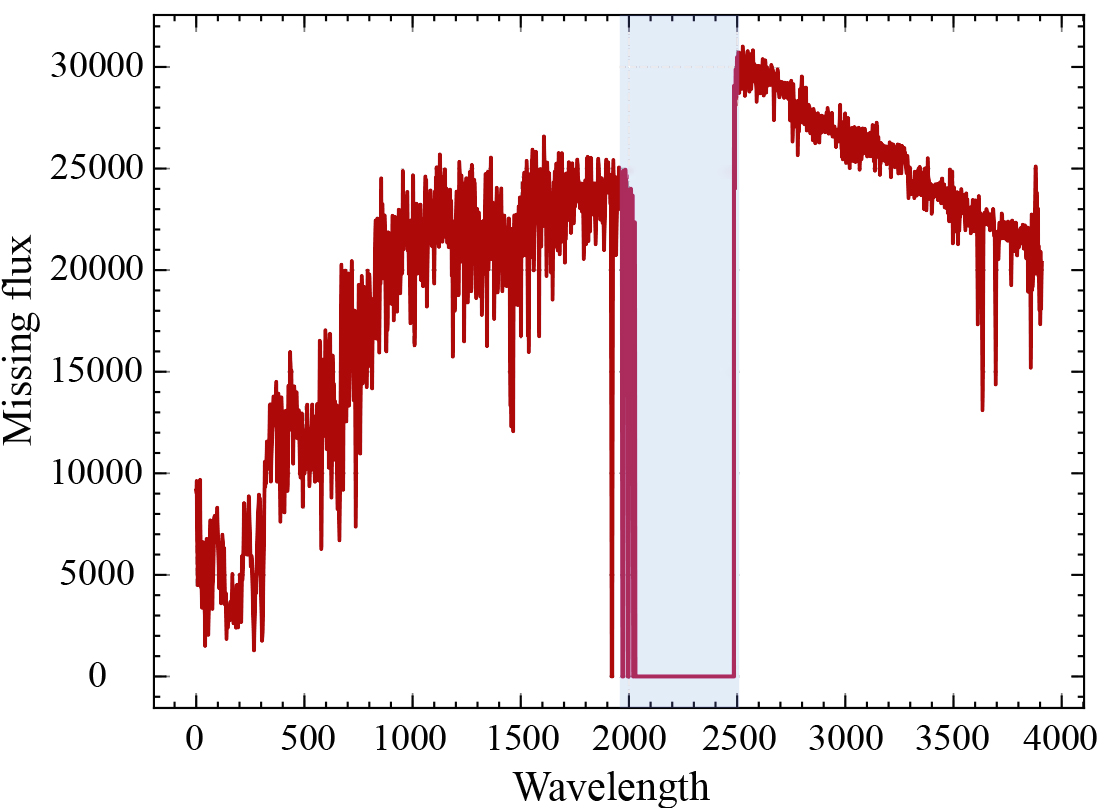}
         \\
         \includegraphics[width=0.46\textwidth]{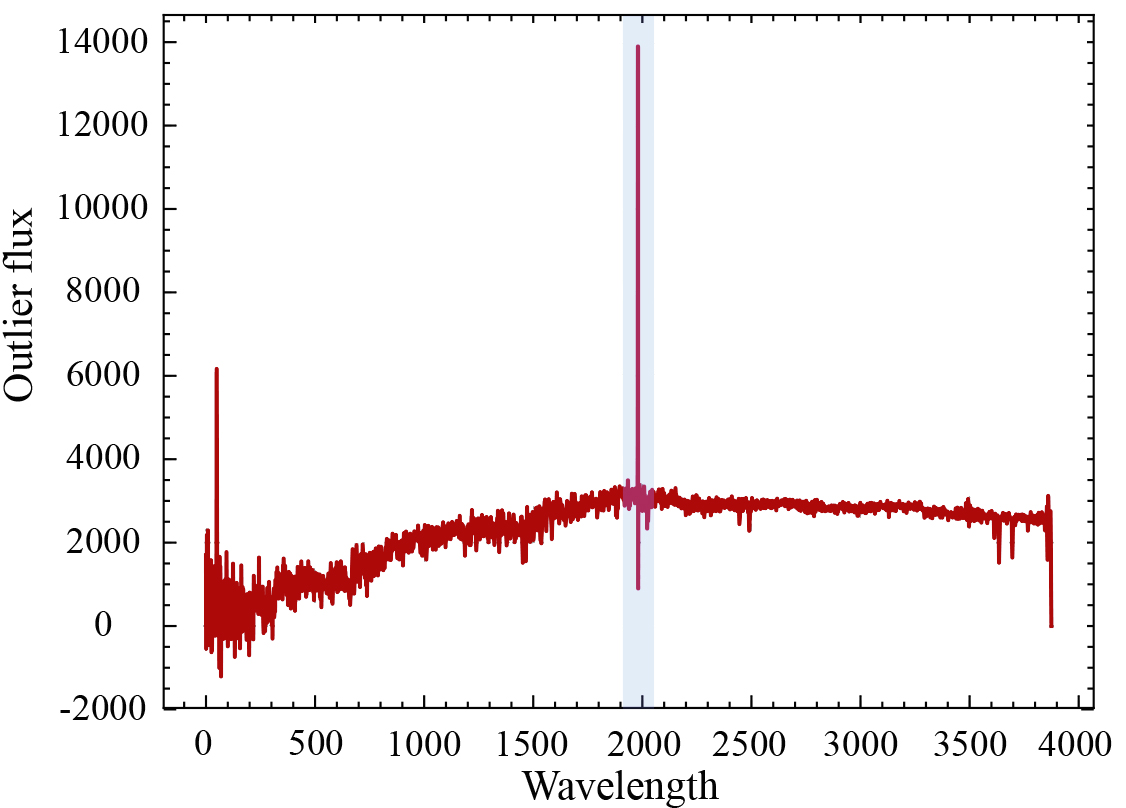}
         \includegraphics[width=0.46\textwidth]{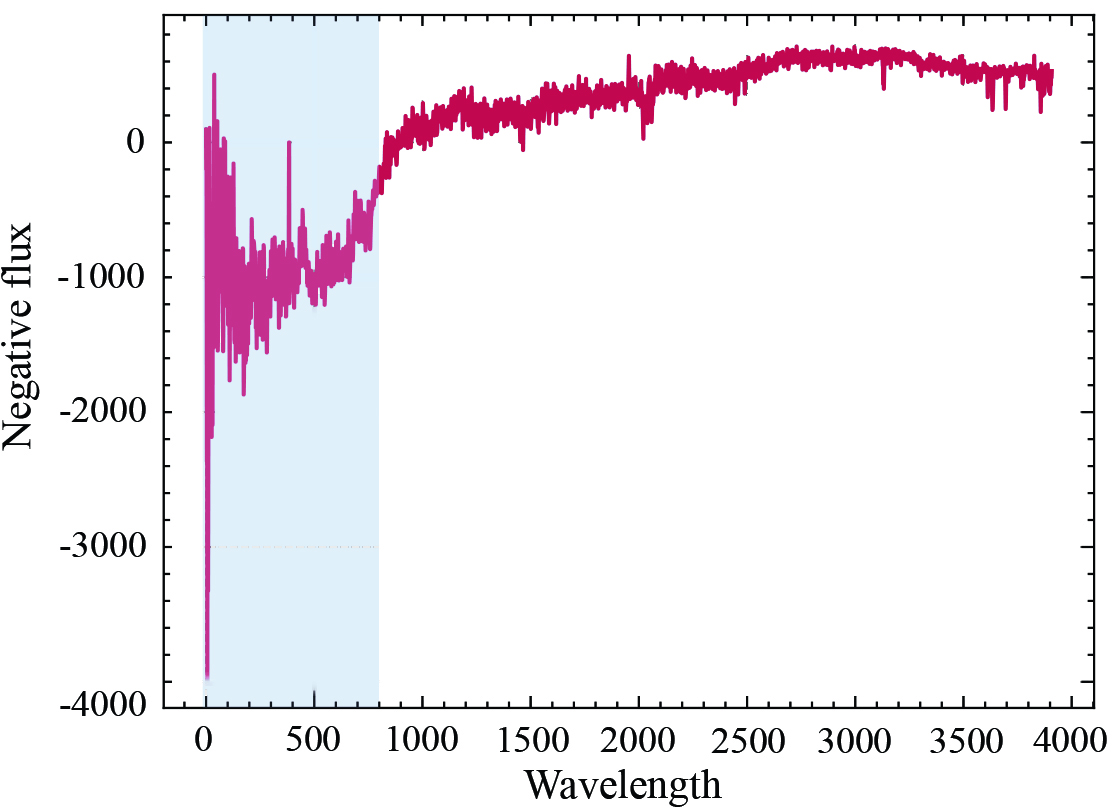}
         
         \caption{Types of anomalous spectra excluded. 
         The red line shows the trend of flux as a function of wavelength. 
         The shaded regions in each panel show the positions of the anomalies. 
         \textit{Top left panel:} Oscillations. The spectrum shows the repetitive variation of flux about a central value. 
         \textit{Top right panel:} Missing values. These spectra lack values in certain wavelength regions. 
         \textit{Bottom left panel:} Outliers. The spectrum has pronounced peaks or deep troughs. 
         \textit{Bottom right panel:} Negative values. The spectra display negative values at certain wavelengths.}
         \label{fig:data_processing}
      \end{figure*}

   \subsection{Monte Carlo Dropout} \label{subsec:MC dropout}
      Monte Carlo Dropout (MC Dropout), as presented by \cite{gal2016dropout}, is a commonly used method to measure the uncertainty 
      in deep learning models. 
      In our research, we have added MC Dropout to our SPT model to asses the confidence level of its predictions and identify regions
      where it might not perform well.

      The core idea of MC Dropout is to preserve the stochastic behavior of dropout layers, even during the prediction phase of deep 
      neural networks. 
      This is achieved by sampling the predictive distribution of the dropout model using Monte Carlo methods, followed by leveraging sampled 
      weights and moment-matching to estimate the predictive mean and uncertainty of the model.

      The mean of the predictive distribution is estimated using the first raw moment, as given by:
      \begin{align}
      \mathbb{E}_{q(\mathbf{y}^*|\mathbf{x}^*)}\left(\mathbf{y}^*\right) \approx \frac{1}{T} \sum_{t=1}^{T} \mathbf{\hat{y}}^* 
      \left( \mathbf{x}^*, W_1^t, \ldots, W_L^t \right)
      \label{eq5}
      \end{align}
      where, $\mathbf{\hat{y}}^* \left( \mathbf{x}^*, W_1^t, \ldots, W_L^t \right)$ represents the output of the model for given input 
      $\mathbf{x}^*$ and weights $ W_1^t, \ldots, W_L^t $,  
      and $q(\mathbf{y}^*|\mathbf{x}^*)$ indicates the distribution of $\mathbf{y}^*$.  
      $T$ indicates the number of forward inferences, with each inference employing a set 
      of instances drawn from the Bernoulli distribution of the weights.

      The estimation of predictive variance is more complex, involving three main components: the inverse model precision term, 
      the second raw moment, and the square of the first raw moment. The specific formula is:
      \begin{equation}
      \begin{aligned}
         &\text{Var}_{q(\mathbf{y}^*|\mathbf{x}^*)}\left(\mathbf{y}^*\right) \approx \tau^{-1} \mathbf{I}_D \\
         &+ \frac{1}{T} \sum_{t=1}^{T} \mathbf{\hat{y}}^* \left( \mathbf{x}^*, W_1^t, \ldots, W_L^t \right)^T 
            \mathbf{\hat{y}}^* \left( \mathbf{x}^*, W_1^t, \ldots, W_L^t \right) \\
         &- \mathbb{E}_{q(\mathbf{y}^*|\mathbf{x}^*)}\left(\mathbf{y}^*\right)^T \mathbb{E}_{q(\mathbf{y}^*|\mathbf{x}^*)}
            \left(\mathbf{y}^*\right)
      \end{aligned}
      \label{eq6}
      \end{equation}
      where $\tau ^{-1}$ represents the inverse of the model's precision and is associated with the weight decay $\lambda $ and 
      prior length-scale $l$, expressed as $\tau^=\frac{pl^2}{2N \lambda}$. 
      $\mathbf{I}_D$ is an identity matrix.
      The inverse model precision term introduces inherent uncertainty within the model, ensuring that the estimated variance does not vanish even 
      when the sample variance approaches zero. 
      Equation. (\ref{eq6}) signifies the model's uncertainty in predictions for a particular input, offering an assessment of the confidence of these
      predictions.

      Through the process described above, MC Dropout not only enables precise predictions of the age and mass of red giants but also 
      quantifies the associated uncertainty. 
      This improvement enhanced the interpretability of the model, helping to understand its behavior across different regions, and potentially 
      identifies areas for investigation and optimization.

   \subsection{Prediction Valuation Metrics} \label{subsec:measure}
      In astrophysical research using regression methods, we typically assess a model based on two criteria: the mean absolute error ($\Delta$) 
      and the mean absolute percentage error ($\Delta_P$).
      
      The $\Delta$ quantifies the average error in the model's predictions, indicating the absolute difference between the predicted and 
      observed values. The $\Delta_p$ measures the relative error, 
      offering information about the percentage discrepancy between 
      the model's predictions and the actual values.
      In formulaic terms, they are given by: 
      \begin{equation}\label{MAE}
         {\Delta}={\frac{1}{n}\sum_{i = 1}^{n}|\hat{y_i}-y_i|}
      \end{equation}

      \begin{equation}\label{MAPE}
         {\Delta_P}={\frac{1}{n}\sum_{i = 1}^{n}\frac{|\hat{y_i}-y_i|}{y_i}\times 100\%}
      \end{equation}
      here, $n$ represents the number of samples in the dataset, while $\hat{y_i}$ and $y_i$ are the predicted and true values for the 
      $i$th sample, respectively.
      These metrics provide an assessment of the model. 
      A combination of low $\Delta$ and $\Delta_P$ values often indicates strong predictive accuracy of the model.

      \begin{figure*}
         \centering
         \includegraphics[width=0.45\textwidth]{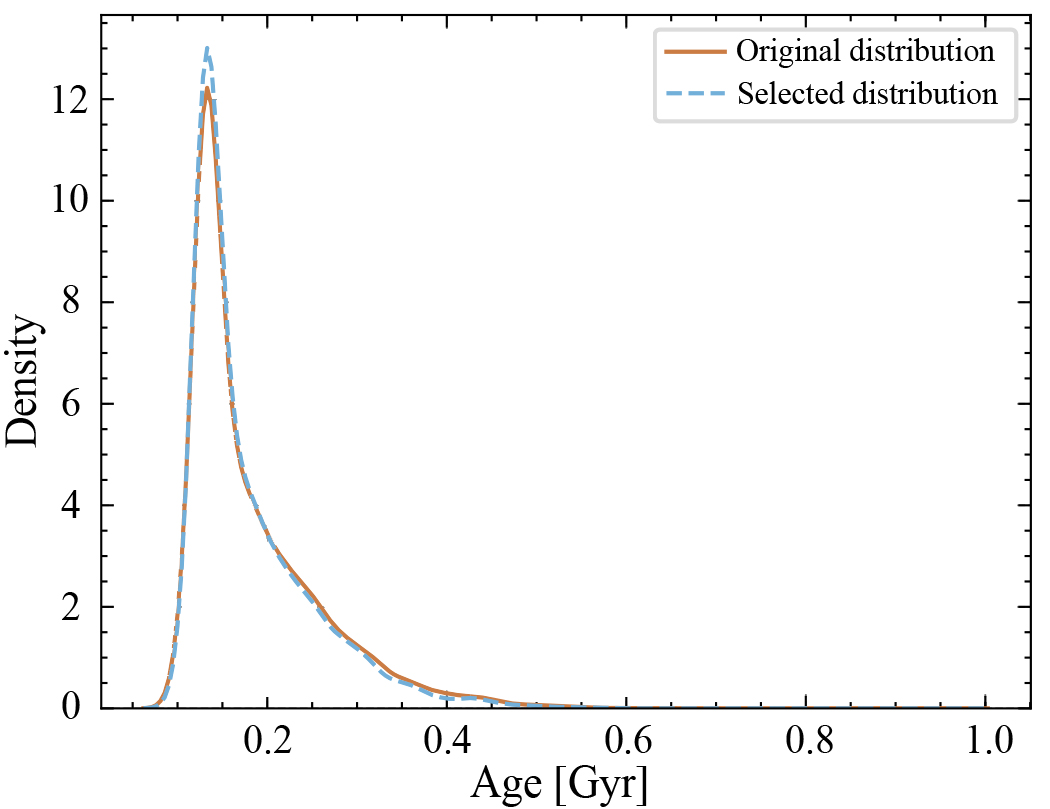}
         \includegraphics[width=0.45\textwidth]{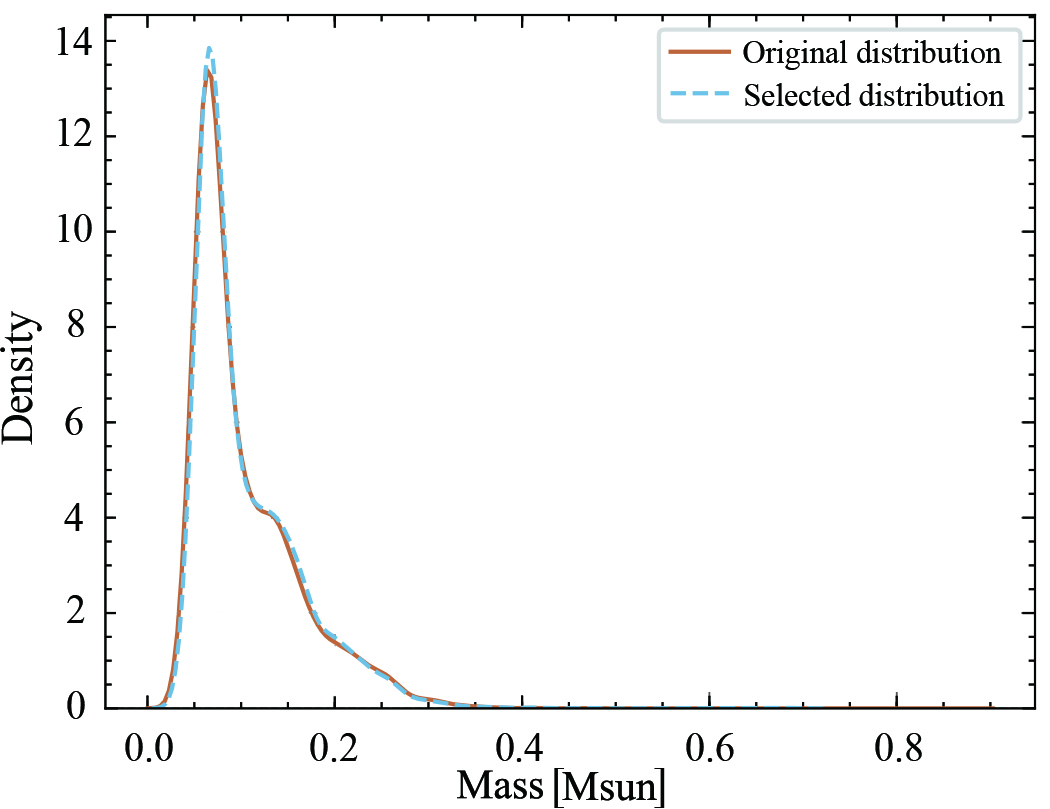}
         
         \caption{
         Pre-and-post data distribution. 
         The solid orange line depicts the original data distribution, while the blue dashed line shows the data distribution after removing outlier spectra. 
         \textit{Left panel:} Changes in the density distribution of age. 
         \textit{Right panel:} Changes in the density distribution of mass.}
         \label{fig:data_distribution}
         \end{figure*}

   \subsection{Training Details} \label{subsec:training details}
   Within the field of deep learning, the selection of optimization algorithms, 
   their hyperparameter configurations, 
   and training tricks affect the performance and results of the model. 
   For optimal efficiency and stability in the SPT, 
   we implemented the following strategies and configurations:
   
   \textbf{Optimization Strategy and Parameters:}\quad The selection of an appropriate optimization algorithm is important to ensure effective 
      model convergence. 
     In this study, we employed the widely used Adam optimizer. 
     This optimizer amalgamates the strengths of both the Momentum and RMSprop algorithms and offers adaptive learning rate adjustments. 
     Empirically, we set the parameters for the Adam optimizer with $\beta_1=0.9$, $\beta_2=0.999$, and $\epsilon=1e^{-8}$ to assure steady 
     model learning.
     
   \textbf{Learning Rate Policy:}\quad We also employed a learning rate decay strategy to further enhance model convergence.
   Specifically, we set the initial learning rate at $1e^{-4}$, and it decreased by a factor of 0.8 after every 100 training epochs.
     
   \textbf{Regularization Technique:}\quad To address potential overfitting, we applied $L_2$ regularization with a coefficient of $5e^{-4}$. 
     This method helps constrain the model weights to converge to smaller values, leading to more consistent and smoother predictions.
   
   \textbf{Weight Initialization Strategy:}\quad We adopted a truncated normal distribution approach for weight initialization.
   
   \textbf{Early Stopping Mechanism:}\quad Specifically, if the model does not show significant improvements on the validation set over ten 
      consecutive training epochs, the training is terminated early.
     In such cases, we use the average of the model weights over the ten epochs as the ultimate model parameters to ensure robustness.

\section{Data} \label{sec:data}
   \subsection{Datasets}\label{dataset}
   Spectroscopy is a powerful tool for investigating the properties of red giants, as we have shown in Sect. \ref{sec:intro}. 
      In this work, we use the spectra from the LAMOST DR9 to 
      estimate the age and mass of red giants. 
      LAMOST is a national facility of China, operated by the National Astronomical Observatories, Chinese Academy of Sciences. 
      It consists of a reflecting Schmidt telescope with 4,000 fibers that can observe 20 $\text{deg}^2$ of the sky simultaneously. 
      As of July 2020, LAMOST has completed its pilot survey, which ran from October 2011 to June 2012, and the first nine years of its regular 
      survey, which started in September 2012 \citep{Abazajian_2004,Zhang_2013,2012RAA....12.1197C,2012RAA....12..723Z}. 
      The survey contains 11,817,430 flux- and wavelength-calibrated, sky-subtracted spectra, of which 11,473,644 are stellar spectra. 
      The spectra cover the wavelength range of 3,700 \text{-} 9,000 \AA \ with a resolution of 1,800 at 5,500 \AA\quad
      \citep{Stoughton_2002,2003AJ....126.2081A}.
      The LAMOST dataset provides a rich source of spectroscopic information for various astrophysical applications. 
      The wide spectral coverage, from ultraviolet to infrared wavelengths, allows us to extract multi-wavelength features that help us better 
      understand the nature of red giants.
   
      \cite{2018ApJS..239...32P} presented an extensive APOKASC\text{-}2 catalog featuring 6,676 evolved red giants, from which ages and masses 
      were derived using state-of-the-art asteroseismology techniques. 
      Their pioneering approach involved a novel empirical method that skillfully integrated results from five independent asteroseismic pipelines.
      Furthermore, they conducted meticulous calibrations of asteroseismic parameters and applied theoretical corrections to the scaling relations 
      of $\Delta\nu $  and the zero-point calibration of $\nu_{\text{max}}$. 
      As a result of these rigorous efforts, highly precise estimates of age and mass were obtained for this cohort of red giants. 
      In more detail, the correct mass $M_{\text{cor}}$  and radius $R_{\text{cor}}$ by
   \begin{equation}\label{Mcor}
      \mathit{\frac{M_{\text{cor}}}{M_{\odot}}}=
     \left({\frac{f_{\nu_{\text{max}}}^3}{f_{\Delta \nu }^4}}\right)
     \left({\frac{\nu_{\text{max}}}{\nu_{\text{max},\odot}}}\right)^{3}
     \left({\frac{T_{\text{eff}}}{T_{\text{eff},\odot}}}\right)^{1.5}
     \left({\frac{\Delta \nu}{\Delta \nu_\odot }}\right)^{-4}
   \end{equation}and
   \begin{equation}\label{Rcor}
      \mathit{\frac{R_{\text{cor}}}{R_{\odot}}}=
     \left({\frac{f_{\nu_{\text{max}}}}{f_{\Delta \nu }^2}}\right)
     \left({\frac{\nu_{\text{max}}}{\nu_{\text{max},\odot}}}\right)
     \left({\frac{T_{\text{eff}}}{T_{\text{eff},\odot}}}\right)^{0.5}
     \left({\frac{\Delta \nu}{\Delta \nu_\odot }}\right)^{-2}
   \end{equation}
   
      They primarily aim to calibrate asteroseismic parameters, $\nu_{\text{max}}^{j}$ and $\Delta \nu^j$, for each red giant star (indexed $j$). 
      \cite{2018ApJS..239...32P} derived scale factors for every pipeline (indexed $i$), denoting them as $X_{\nu_{\text{max}}}^i$ 
      and $X_{\Delta \nu}^i$. 
      This strategy ensures uniform average results across all pipelines within the sample set. 
      Once these scale factors are in place, measurements are adjusted and averaged to determine asteroseismic parameters for each red giant. 
      Importantly, the scale factor for a pipeline represents a weighted average based on the dispersion of absolute measurements, providing stars 
      with larger dispersions with proportionally reduced weights.
      
      Given the impact of systematic errors, the scaling relation for $\Delta \nu$, grounded in theory, has been revised. 
      Moreover, the scaling relation for $\nu_{\text{max}}$ underwent calibration within the context of open clusters. 
      Consequently, the initial assumption of $f_{\nu} = f_{\nu_{\text{max}}} = 1$ has been superseded by the derived $f_{\nu}$ and 
      $f_{\nu_{\text{max}}}$ values.
      
      Upon obtaining precise mass measurements, the age of each red giant star is determined using interpolation based on the BeSPP grid. 
      In summary, this high-quality age and mass data serve as ideal samples for our deep learning network. 
      Subsequently, a cross-match was conducted between the 6,676 red giant stars and the LAMOST DR9 dataset, yielding a total of 5,057 corresponding
      spectra.
   
      \begin{figure}[!h]
         \begin{center}
          \includegraphics[width=\linewidth]{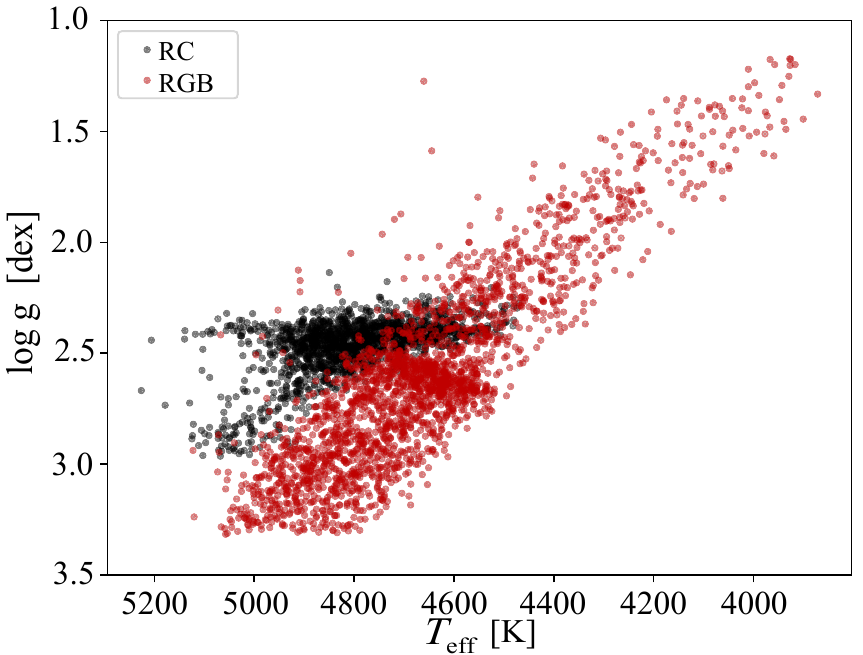}  
         \end{center}
         \caption{
         Distribution of RGB and RC stars in the $T_{\text{eff}}$ - $\log g$ plane.  
         Red points are RGB stars, and black points are RC stars.
         \label{fig:RGB_RC}}
       \end{figure}
       
   \subsection{Data Processing} \label{sec:data_processing}
      When conducting spectroscopic observations on individual stars, LAMOST measures flux across various wavelengths. 
      To ensure uniformity in our input samples, we applied linear interpolation to the spectra spanning the wavelength range of 3,800 \text{-} 9,100 
      \AA. 
      This process yielded a spectral dimension of 4,096.
      
      Given the anomalous spectra present in the dataset, we initiated a data cleaning procedure. 
      To ensure the model's generalization capacity within real sample spaces, we avoided sample selection based on relative age errors. 
      Our aim was to curate high-quality samples by filtering out spectra with irregularities such as oscillations, missing values, outliers, and 
      negative readings, which are presented in the Fig. \ref{fig:data_processing}. 
      Furthermore, the distributions of the age and mass , both pre- and post-sample selection, are shown in Fig.
      \ref{fig:data_distribution}. 
      The peak positions and distribution widths for both age and mass remain consistent before and after the exclusion of anomalous spectra.
      This indicates that the central tendency and variability of the dataset were retained during the cleaning process, underscoring the 
      effectiveness of the cleaning procedure, confirming consistent distributions.
      Finally, a total of 3,880 samples were retained for our study.
 
      We adopted the Maximum Overlap Discrete Wavelet Transform (MODWT) model \citep{chernick2001wavelet} to reduce spectral noise. 
      Especially effective for sequence analysis \citep{2020arXiv200209535W}, 
      MODWT not only offers enhanced computational efficiency over the Continuous Wavelet Transform 
      but also provides translation invariance, a feature not present in the Discrete Wavelet Transform.
      The Haar wavelet served as our typical mother wavelet choice. 
      
      To minimize data collinearity and strengthen regression model stability, we applied PCA for dimensionality reduction 
      \citep{1994VA.....38..331S,2014NewA...28...35B,2022AJ....163..153L}. 
      We integrated the top 143 principal components, covering 99.9\% of cumulative variance, into our model.

\begin{figure}
     \centering
      \includegraphics[width=\linewidth]{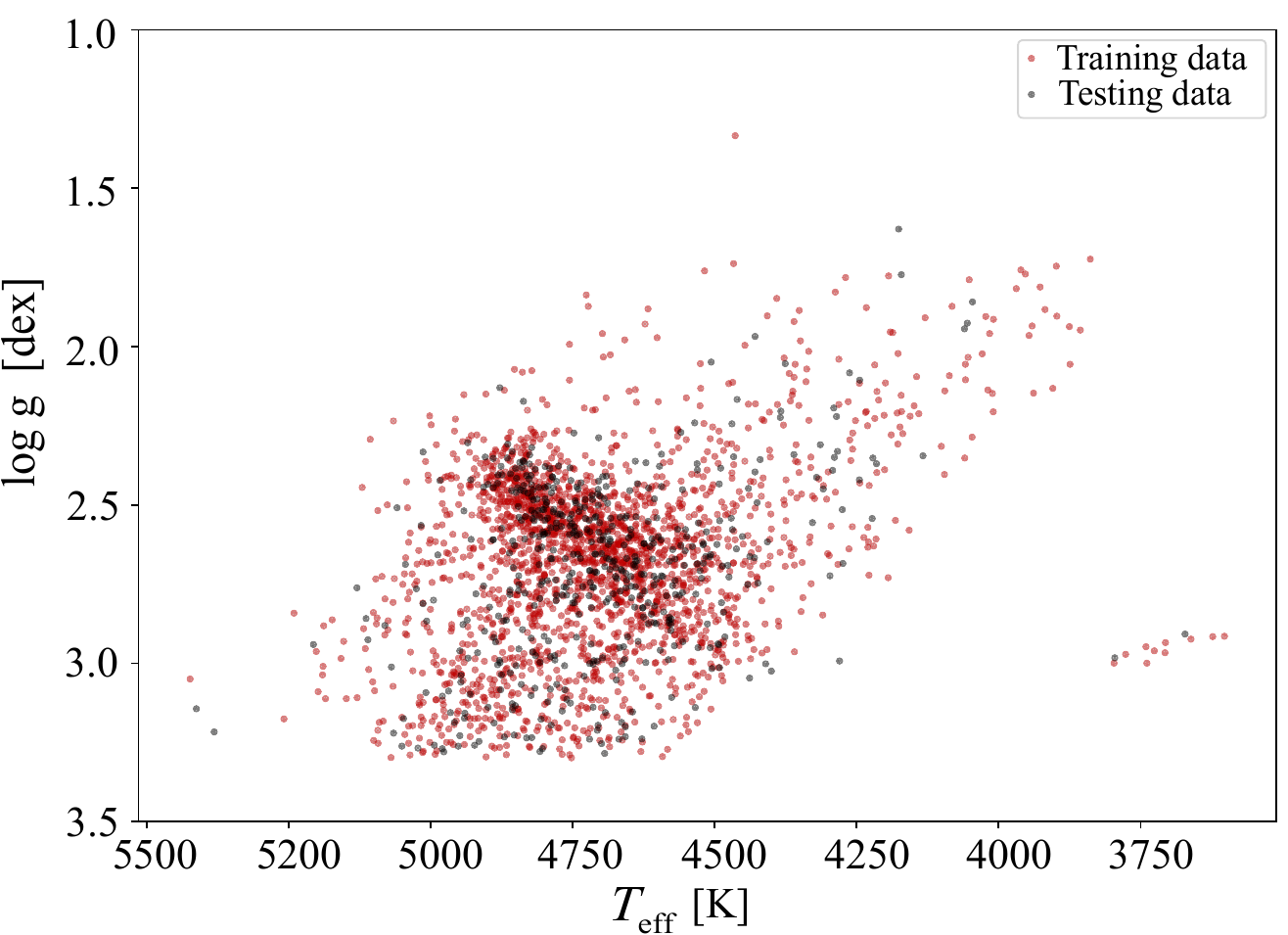}  
     \caption{Partitioning of Samples. 
     The scatter plot illustrates the data distribution before and after sample division, with red points representing the training set and black points representing the test set.}
     \label{fig:HRD}
   \end{figure}

      Figure \ref{fig:RGB_RC} distinctly depicts our final sample of 3,880 red giant stars, 59.7\% are categorized as RGB (H-shell burning),
      while 40.3\% belong to the red clump (RC, or He-core burning) classification \citep{montalban2013testing}.  
      For more details about the stars, please refer to \cite{2018ApJS..239...32P}.
      To evaluate our SPT model, we divided our dataset into training and testing sets in an 8 : 2 ratio. 
      Figure \ref{fig:HRD} demonstrates that both are independently and uniformly distributed. 
      This approach ensures a rigorous and fair evaluation of predictive capabilities of SPT. 
   
   \begin{figure}
      \centering
       \includegraphics[width=\linewidth]{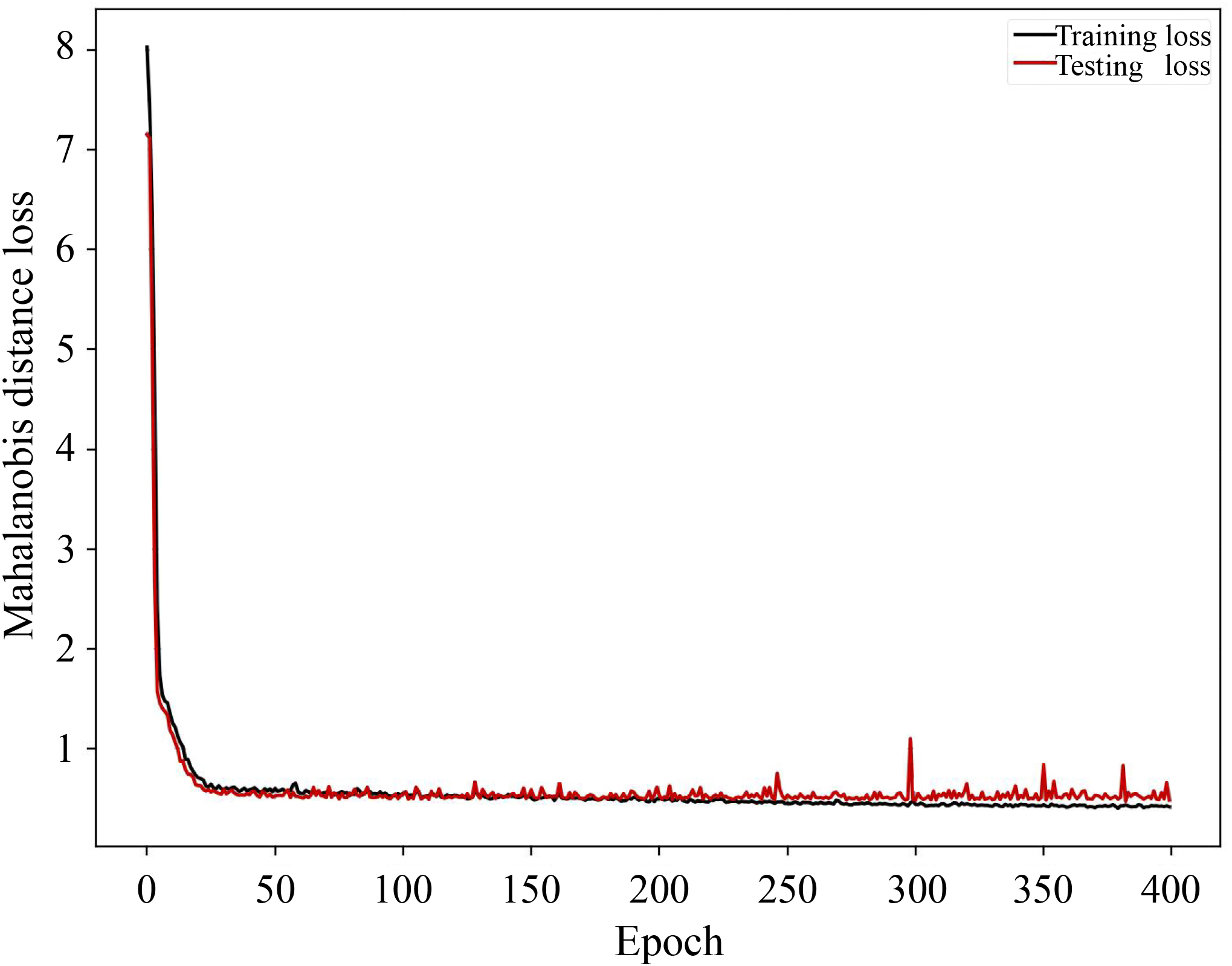}  
      \caption{Learning curve based on Mahalanobis distance loss function. 
      This graph illustrates the changes in the loss value based on Mahalanobis distance. 
      The black and red curves represent the variations in loss values on the training and testing sets, respectively.}
      \label{fig:learning curve}
    \end{figure}

    \section{Results} \label{sec:result}
    Utilizing the strategies and configurations detailed in Sect. \ref{subsec:training details}, our model not only demonstrated superior 
    performance but also ensured rapid and consistent convergence.
    This forms a reliable basis for accurate determinations of the ages and masses of red giants. 
    Figure \ref{fig:learning curve} offers a detailed insight into the training process of our model. 
    It shows that the performance of the model has gradually converged on both the training and validation sets, highlighting the stability 
    of the training process and the generalizability of the model.

   \subsection{Age and Mass Estimation} \label{subsec:age estimation}
     
   \begin{figure*}
      \centering
      \includegraphics[width=0.45\textwidth]{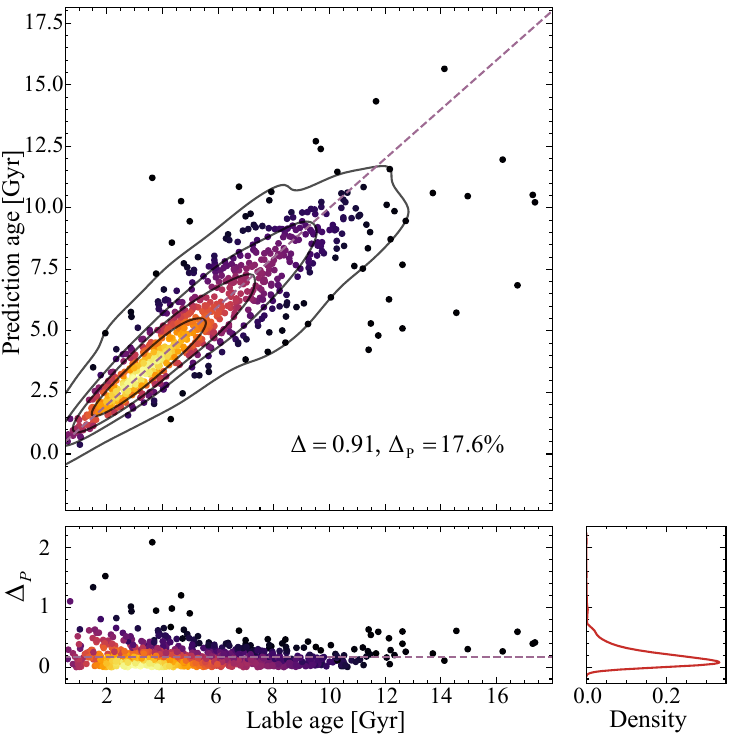}
      \includegraphics[width=0.45\textwidth]{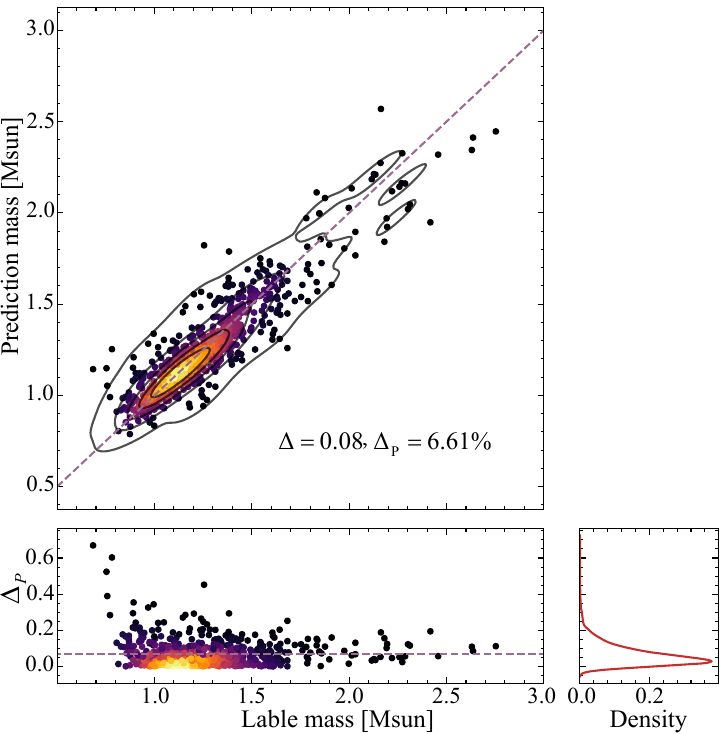}
      \caption{
      Regression results for age and mass of red giants. 
      The upper sections of each panel display kernel density plots of predicted values versus true labels, where a yellower color indicates higher 
      density, signifying a more concentrated data; 
      the dashed lines represent the identity line, while the black lines indicate isohypse. 
      $\Delta$ and $\Delta_P$ represent the mean absolute error and mean absolute percentage error, respectively. 
      The lower sections show scatter plots and distribution graphs for $\Delta_P$ (with the red line). 
      The scatter plots demonstrate the error of the predictive model at each data point, with the color-density relationship consistent with the upper panels; 
      the dashed lines denote the average value of $\Delta_P$, providing a reference for the overall error level. 
      The distribution graphs illustrate the spread of $\Delta_P$ across the entire dataset. 
      “Density” is the probability density fitted to the $\Delta_P$.
      \textit{Left panel:} Regression results for ages with an average absolute error $\Delta$ of 0.91 dex and $\Delta_P$ of 17.64\%. 
      \textit{Right panel:} Regression results for masses with $\Delta$ and $\Delta_P$ values of 0.08 dex and 6.61\%, respectively.}
      \label{fig:regression_results}
   \end{figure*}
   
   In our study, we employed the SPT to process 3,880 spectra. Compared to previous age estimation works based on machine learning 
   methods \citep{2019MNRAS.484..294D, 10.1093/mnras/stad1272}, 
   our method does not necessitate the pre-extraction of explicit high-precision parameters from the spectra, 
   such as \(T_{\text{eff}}\), \(\log g\), and \(\text{[Fe/H]}\). 
   This allows our approach to be free from dependence on prior spectroscopic data preprocessing. 
   Furthermore, due to the data-driven nature of machine learning, the richer spectral information allows us to 
   mitigate the inductive bias that might arise from characterizing spectra solely based on spectroscopic parameters.

   The left panel of Fig. \ref{fig:regression_results} clearly depicts well-estimated and robust predictions, 
   indicating \(\Delta\) of 0.91 dex and \(\Delta_P\) of 17.64\%. While 6\% of the predictions exceeded 50\%, 
   the majority were within acceptable bounds. However, the model encountered difficulties when estimating ages close to the cosmic limit of 13.8 Gyr, 
   likely due to inaccuracies in the asteroseismic parameters.

   As shown in the right panel of Fig. \ref{fig:regression_results}, 
   we evaluated the estimated masses across our red giant sample, analyzing the distribution of their errors. 
   The majority of stars have mass estimates ranging from 0.9 to 1.4 \(M_{\odot}\), which aligns well with true labels. 
   The validation results in \(\Delta\) of 0.08 dex and \(\Delta_P\) of 6.61\%. Although 1.8\% of the red giants exhibited
   notably large errors exceeding 40\%, these outliers might be associated with stars in distinct evolutionary phases or 
   affected by unaccounted astronomical processes. Nevertheless, the majority of our mass estimations lie within a satisfactory range.

    \begin{table}
      \caption{Comparison with Other Machine Learning Methods}
      \label{tab:method}
      \centering
      \begin{tabular}{lllll}
      \hline \hline
      Model & \multicolumn{2}{c}{Age Results} & \multicolumn{2}{c}{Mass Results} \\
      \cline{2-3} \cline{4-5}
      & \multicolumn{1}{c}{$\Delta$} & \multicolumn{1}{c}{$\Delta_P$} & \multicolumn{1}{c}{$\Delta$} & \multicolumn{1}{c}{$\Delta_P$} \\
      \hline
      RF & 1.40 & 36.03\% & 0.14 & 10.78\% \\
      XGBoost & 1.20 & 28.31\% & 0.12 & 9.37\% \\
      LightGBM & 1.22 & 27.77\% & 0.11 & 9.04\% \\
      CatBoost & 1.11 & 25.10\% & 0.11 & 8.16\% \\
      Ours & 0.91 & 17.64\% & 0.08 & 6.61\% \\
      \hline
      \end{tabular}
   \end{table}

   \subsection{Uncertainty Analysis} \label{subsec:uncertainty analysis} 

      \begin{figure*}
         \raggedright
         \includegraphics[width=0.45\textwidth]{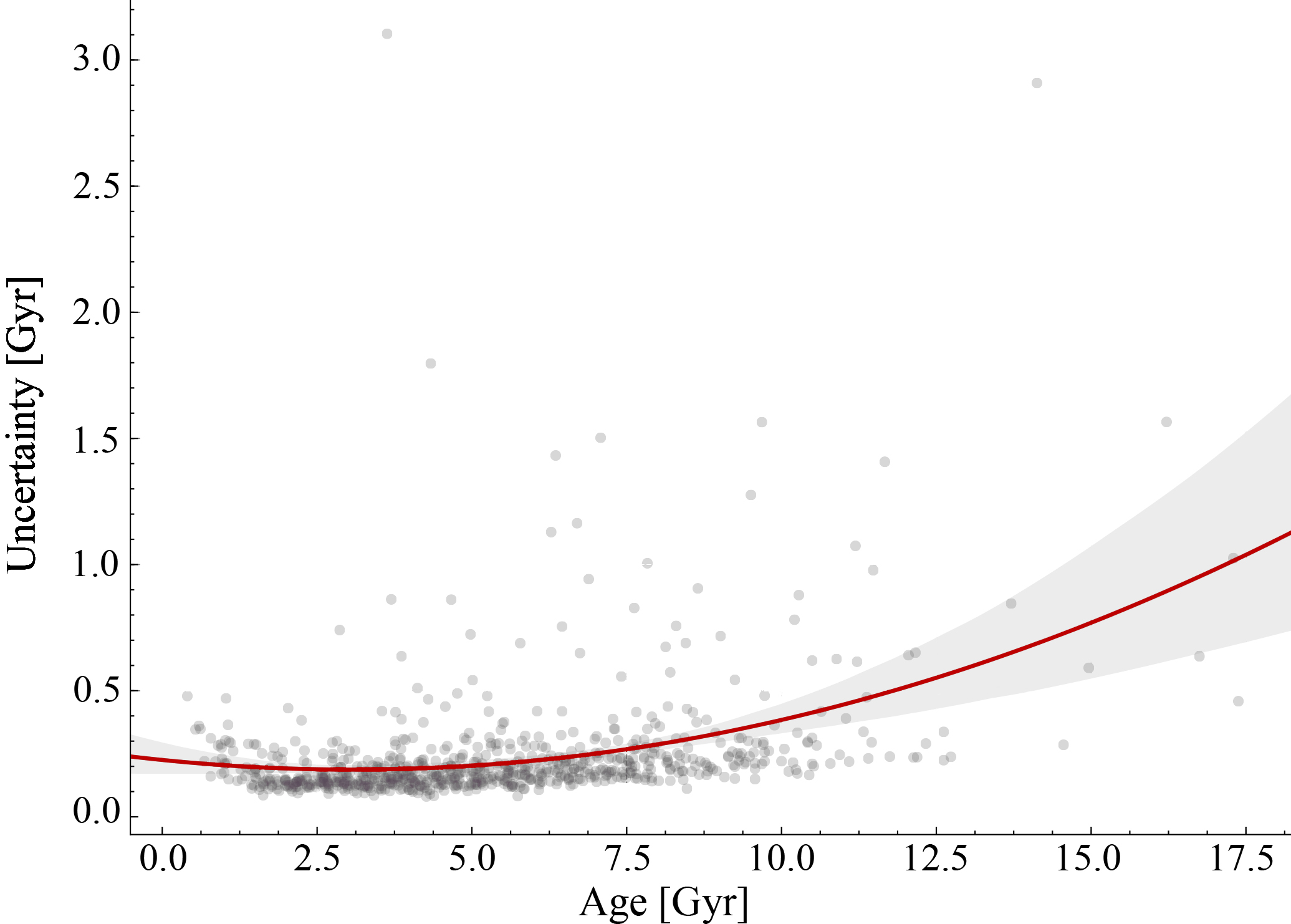}
         \includegraphics[width=0.45\textwidth]{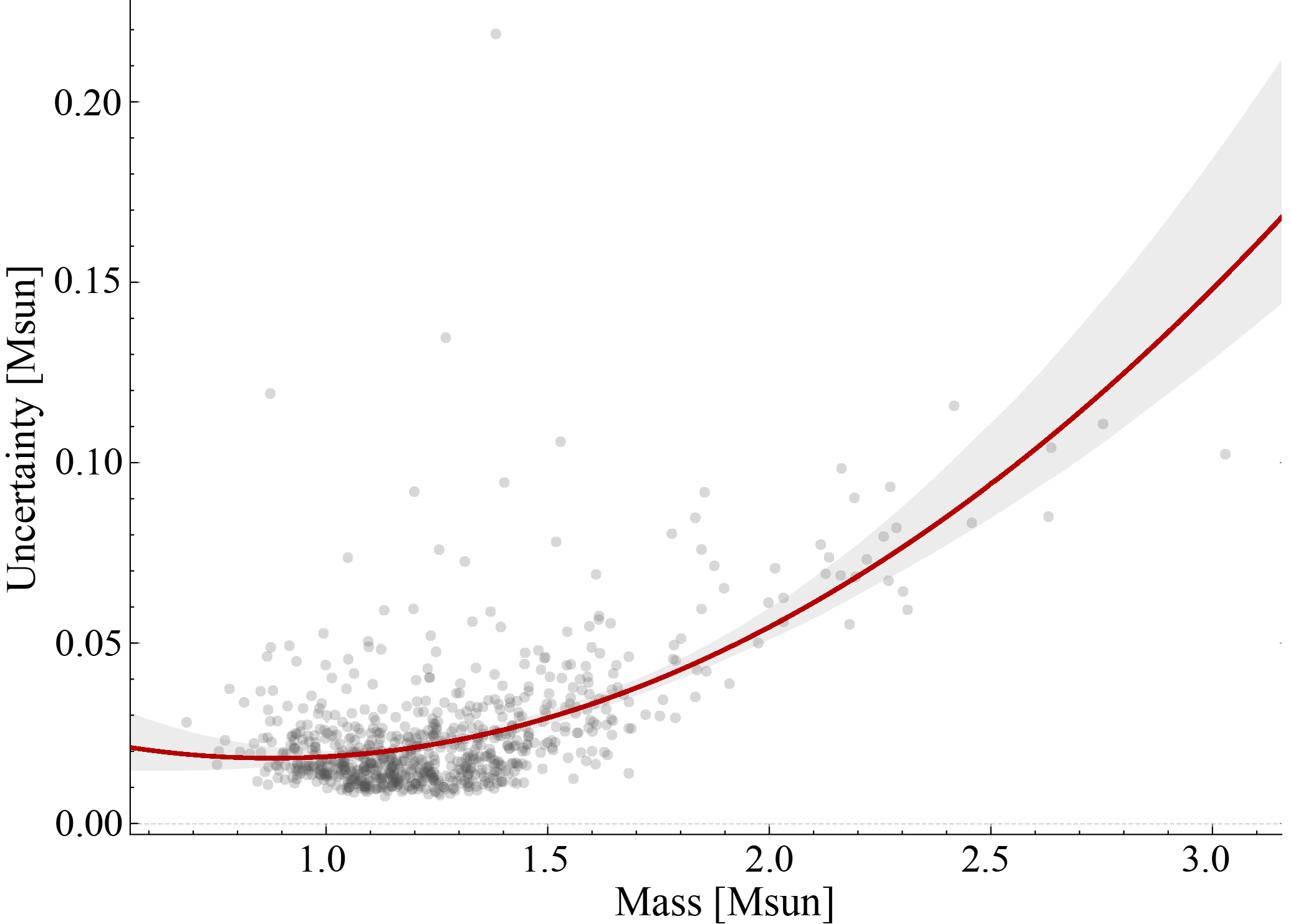}
         \caption{
         Uncertainties for age (\textit{left panel}) and mass (\textit{right panel}) predictions. 
         The red solid lines indicate the trend of uncertainties relative to the true labels. 
         The shaded regions correspond to the 95\% uncertainty intervals, representing the primary range of 
         uncertainties provided by the model.}
         \label{fig:uncertainties}
      \end{figure*}

      \begin{table}
         \caption{Comparison of Age Predictions for Red Giants in Open Clusters with Literature Values\label{tab:open_cluster}}
         \centering
         \begin{tabular}{l l l l l}
             \hline \hline
             Data & Cluster & $\text{Age}_{\text{Liter}}$ & $\text{Age}_{\text{Ours}}$ & $\text{Number}_\text{{Star}}$ \\
             & & [Gyr] & [Gyr] & \\
             \hline
             LAMOST & M 67 & 4.0 & 4.0 & 36\\ 
                    & Berkeley 32 & 6.0 & 6.0 & 10 \\
             \hline
             APOGEE & FSR 1077 & 0.86 & 1.40 & 3\\
                    & NGC 2158 & 2.14 & 2.39 & 6\\
                    & NGC 2682 & 3.43 & 3.41 & 28\\
             \hline
         \end{tabular}
     \end{table}

  To rigorously assess prediction reliability, 
  we incorporated MC dropout for uncertainty quantification and illustrated 
  the relationship between its uncertainties and the age and mass estimations 
  of red giants in Fig. \ref{fig:uncertainties}.

  We observed a specific distribution trend: lower uncertainties are manifested in the central region of the prediction range, 
  where samples are more abundant. In contrast, peripheral areas show increased uncertainties. 
  This pattern suggests that in the central data region, the presence of potentially more conventional training samples 
  provides the SPT with more remarkable stability and higher confidence, effectively capturing primary data trends and patterns. 
  This ensures the reliability of the SPT in most scenarios. Conversely, when predictions approach the boundaries of its training data,  
  the SPT becomes more cautious due to factors like fewer samples and distinct data characteristics from the center, 
  exhibiting heightened sensitivity against rare or atypical situations. Such discernment indicates that the SPT can adopt a more judicious stance, 
  alerting users about its current predictions when confronted with unfamiliar data, rather than making potentially misleading blind predictions.

\section{Validation} \label{sec:valuation}
   \subsection{Machine Learning Algorithms versus SPT} \label{subsec:model comparisons}
      As machine learning has grown, many algorithms have been used for different tasks.
      Notably, the estimation of age and mass in red giants remains a challenge. 
      In our study, we compare our method to several machine algorithms including Random Forest \cite[RF;][]{breiman2001random}, XGBoost 
      \citep{2016arXiv160302754C}, LightGBM \citep{ke2017lightgbm}, and CatBoost \citep{2018arXiv181011363V}. 

   \begin{figure*}
      \centering
      \includegraphics[width=0.45\textwidth]{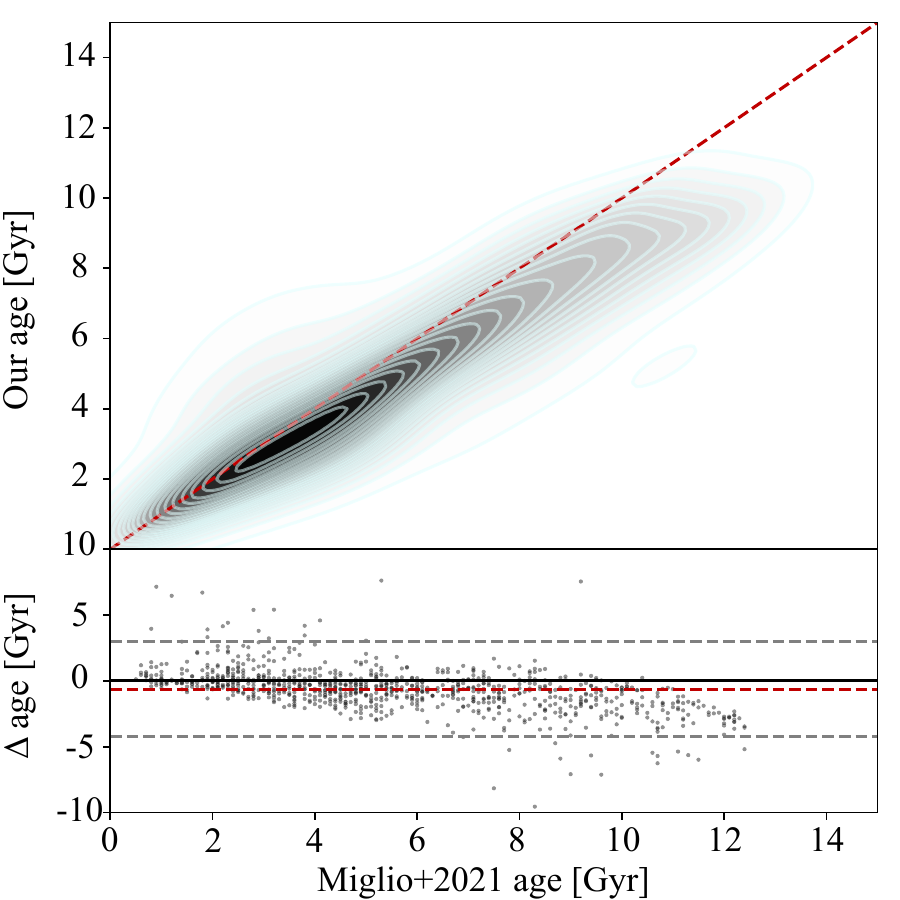}
      \includegraphics[width=0.45\textwidth]{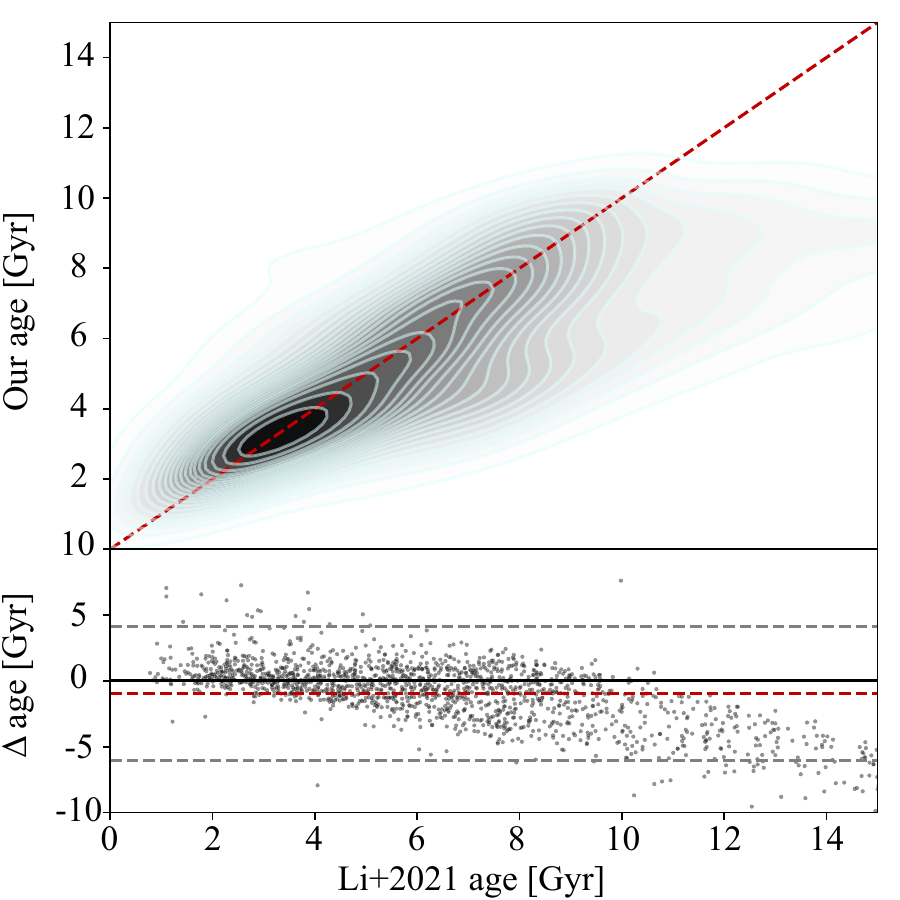}
      
      \includegraphics[width=0.45\textwidth]{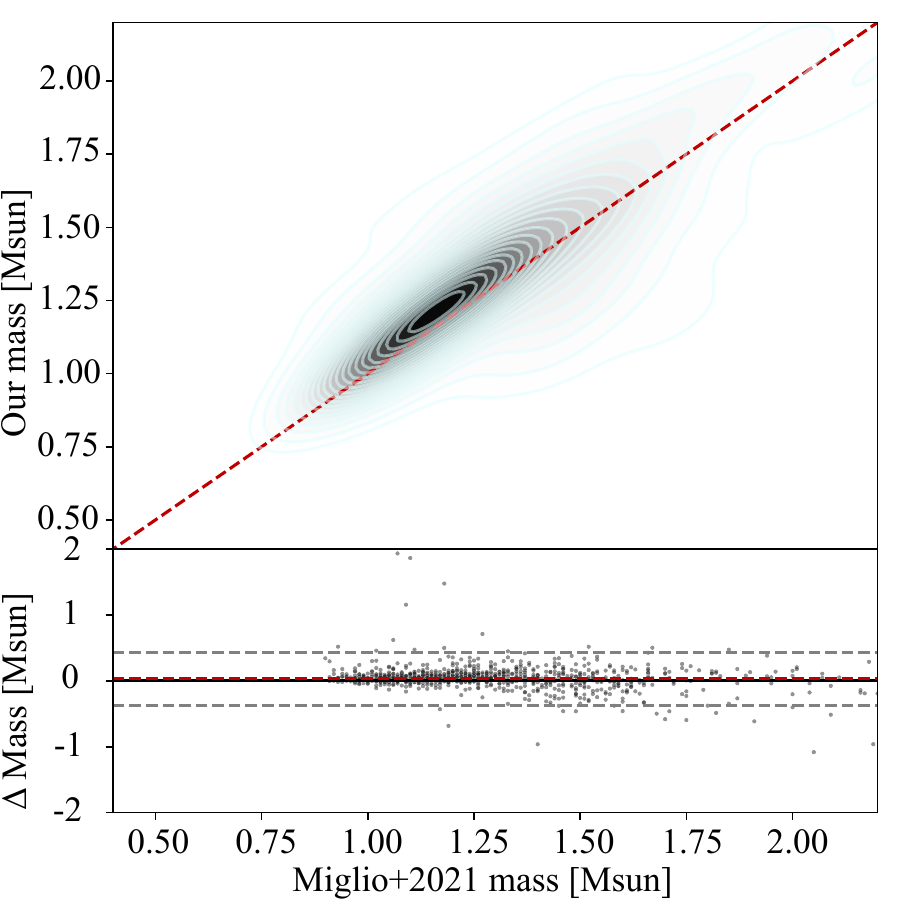}
      \includegraphics[width=0.45\textwidth]{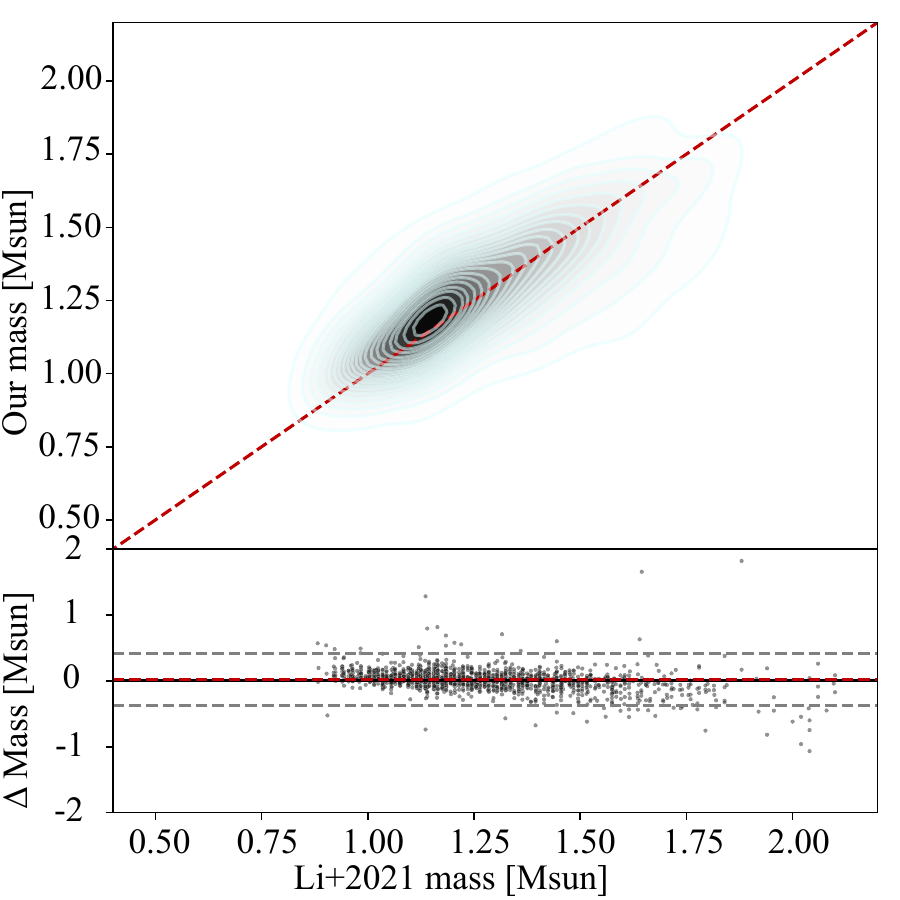}
      
      \caption{
      Comparison of age and mass predictions for red giants using various asteroseismic methods. 
      \textit{Top left panel:} Comparison of age predictions with \cite{2021A&A...645A..85M}. 
      \textit{Top right panel:} Comparison of age predictions with \cite{2022ApJ...927..167L}. 
      \textit{Bottom panels:} Mass predictions comparisons akin to top panels. 
      The upper sections of each panel display kernel density plots related to age or mass as compared with asteroseismic methods, 
      with darker colors indicating higher densities;
      red dashed lines represent the identity line, while the light cyan lines indicate isohypse.
      The lower sections depict residuals, with gray points showing the deviations between our model and the asteroseismic benchmarks. 
      A horizontal red dashed line indicates the mean residual, serving as an indicator of overall bias. 
      Two gray dashed lines outline the 95\% consistency boundaries.}
      \label{fig:comparison with asteroseismology}
   \end{figure*}
     
      Table \ref{tab:method} presents a performance comparison of thses algorithms. 
      Our SPT model outperforms the others, achieving the lowest errors in both age and mass predictions. 
      Specifically, compared to the second-best method, CatBoost, our algorithm reduces $\Delta_P$ by nearly 7.5\% for age and 1.5\% for mass.

   \subsection{Comparison with Asteroseismology} \label{subsec:asteroseismology}
      Asteroseismology examines the oscillations caused by acoustic waves within stars. 
      These oscillations provide deep insights into the internal structures of stars, revealing their evolutionary stages, ages, and masses. 
      Given the widespread use of asteroseismic methods to determine stellar age and mass, we aim to compare our SPT 
      with existing asteroseismic methods to evaluate its accuracy and effectiveness.

      To this end, we selected two datasets from the works of \cite{2021A&A...645A..85M} and \cite{2022ApJ...927..167L}. 
      The former, \cite{2021A&A...645A..85M}, utilized Kepler data to analyze the ages of red giants in the Milky Way. 
      Meanwhile, \cite{2022ApJ...927..167L} focused on refining asteroseismic scaling relations, through detailed model adjustments, 
      incorporating complex stellar phenomena like nuclear reactions, convection, and radiation. 
      Given that these methods are based on distinct assumptions and data processing techniques, benchmarking against them allows for a thorough 
      assessment of the accuracy and generalizability of our model.

      Both datasets were preprocessed to fit the SPT model, 
      through spectral linear interpolation, noise reduction, normalization, 
      and dimensionality reduction.
      Figure \ref{fig:comparison with asteroseismology} illustrates the comparison between the predictions of the SPT model and the datasets from
      \cite{2021A&A...645A..85M} and \cite{2022ApJ...927..167L}. 
      The majority of data clustering near the diagonal line indicates a consistency between the SPT predictions and conventional asteroseismic 
      methods.

      Furthermore, Fig. \ref{fig:comparison with asteroseismology} displays scatter plots beneath each subplot, highlighting the differences 
      between the methods. 
      Combining statistical analysis with the scatter plot observations, we perform a robust evaluation of the alignment of the SPT method with 
      established asteroseismic approaches.
      The vertical axis of these plots displays the differences in prediction, represented by gray dots. 
      The mean difference '$\overline{d} $' (represented by the red dashed line) is used to estimate the bias between the two methods.
      The variation of $\overline{d} $ is described by the standard deviation of the differences, denoted as $S_d$.
      The range $\left[\overline{d}-1.96S_{d}, \overline{d}+1.96S_{d}\right] $ outlines the 95\% Limits of Agreement (LoA), highlighted by two 
      gray dashed lines. 
      With most gray points within the LoA and centering around zero, suggests a alignment that the SPT method is in alignment with 
      traditional approaches. 
      Notably, the proximity of the red dashed line to the zero-difference solid black line in the mass estimation graph indicates consistency 
      between the techniques. 
      In conclusion, our results highlight the reliability of our model across various datasets, confirming its effectiveness in estimating 
      red giant ages and masses.

      \begin{figure*}
         \centering
         \includegraphics[width=0.45\textwidth]{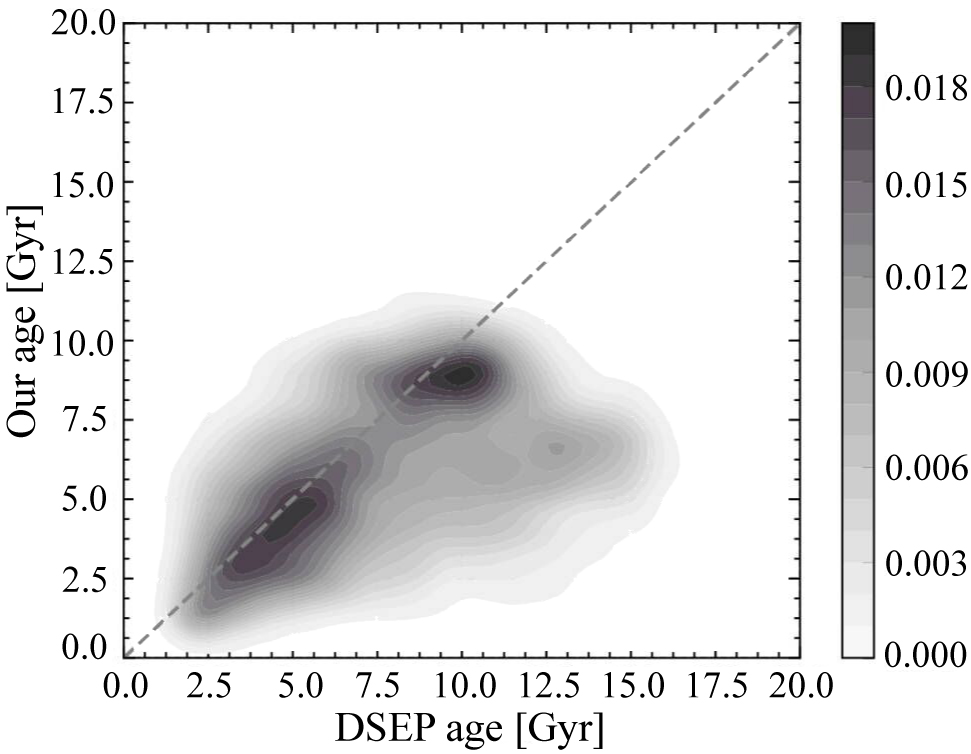}
         \includegraphics[width=0.45\textwidth]{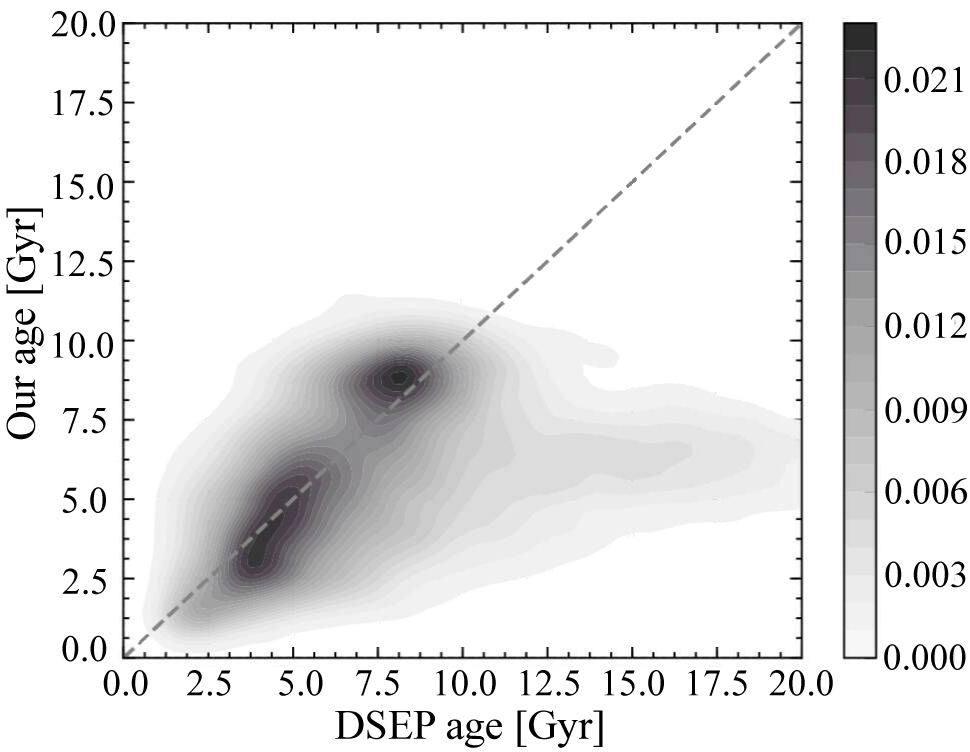}
         \caption{Comparison of ages with different isochrone methods. 
         The horizontal axis represents the age predictions from different isochrone methods, while the vertical axis shows the predictions 
         from our model. 
         \textit{Left panel:} Results in comparison with DSEP \citep{Dotter_2008}. 
         \textit{Right panel:} Comparison with MIST \citep{Choi_2016}.}
         \label{fig:isochrone}
     \end{figure*}

   \subsection{Isochrone Fitting: A Benchmarking Exercise} \label{subsec:isochrones}
      Although isochrone fitting methods have displayed certain constraints in predicting the ages of red giants as mentioned in Sect. \ref{sec:intro}, their robust theoretical 
      underpinnings and demonstrated stability over extended applications solidify their essential role in the astronomical 
      discipline. 
      Traditionally, such methods have served as foundational tools in stellar physics, serving as a widely accepted and familiar benchmark. 
      Given this context, we specifically compared the predictions of our SPT model to those derived from two prevalent isochrone methods, the 
      Dartmouth Stellar Evolution Program \cite[DSEP;][]{Dotter_2008} and the MESA Isochrones and Stellar Tracks \cite[MIST;][]{Choi_2016}.

      From the results shown in Fig. \ref{fig:isochrone}, we observe that while outliers are present, a majority data points cluster around 
      the diagonal, indicating that in most instances, predictions from our SPT model align consistently with traditional isochrone methods.

   \subsection{Red Giant Analysis within Open Clusters} \label{subsec:cluster}
      Open clusters are ideal environments for the study of stellar ages, given that their member stars are typically understood to have a 
      shared origin and age \citep{2019A&A...627A..35C, 2022A&A...660A...4H, 2023MNRAS.518.1505K}. 
      In light of this, we used red giants within open clusters as standard references to validate the age-predictive capabilities of our
      SPT.

      We sourced data from LAMOST DR4 to identify red giant members within the open clusters M 67 and Berkeley 32. 
      Similarly, we used the APOGEE dataset to determine red giant members in the open clusters FSR 1077, NGC 2158, and NGC 2682 
      \citep{2020ApJ...903...55P}.

      Table \ref{tab:open_cluster} presents the ages derived from the literature, predictions from our model. 
      Remarkably, the age predictions of our SPT model for the clusters M67 and NGC 2682 are consistent with the values from the literature, 
      highlighting the reliability of the method. 
      The age predictions for Berkeley 32 align with the documented ages. 
      However, a noticeable age disparity was evident in FSR 1077, but given its limited star sample size, such a variance remains statistically
      acceptable. 

      \setlength{\parskip}{3pt}
      In summary, the age estimates derived from the SPT method consistent with literature values.  
      This confirms not only the theoretical superiority of our model but also its practical accuracy and robustness in application.

\section{Conclusion} \label{sec:conclusion}
   This study successfully developed a deep learning model based on SPT to provide an integrated framework for the 
   simultaneous regression of age and mass directly from red giant spectra. 
   Our SPT employs a novel attention mechanism, the Multi-head HSA operator, which combines the global focus of self-attention with the 
   multi-perspectives observation of multi-head attention. 
   This enables the efficient processing of continuous high-dimensional spectra and overcomes the limitations of CNNs and RNNs in capturing 
   long-distance dependencies.
   We also mathematically substantiated the theoretical efficacy of this operator. 
   Furthermore, we introduced a loss function based on Mahalanobis distance to address age and mass scale imbalances and the loss of 
   interactive modes, enhancing prediction accuracy and robustness. 
   The combination of MC dropout technology allowed for the quantification of uncertainty through sampling weights and moment-matching, 
   providing confidence in specific predictions.

   Our SPT model achieved outstanding estimation accuracy for the age and mass of red giants, with average percentage errors 
   $\Delta_P$ of 17.64\% and 6.61\%, respectively.
   It also provides confidence levels for each prediction. 
   This assists in alerting misguided predictions when encountering unfamiliar data. 
   Additionally, the SPT model demonstrated not only significant superiority over traditional machine learning algorithms but also showed remarkable 
   consistency when compared with conventional asteroseismology methods and isochrone fitting techniques. We also validate the reliability of the predictions in 
   open clusters.
   These findings emphasize the superior performance of the SPT model, offering 
   new perspectives and tools for the estimation of age and mass of red giant.

   The upcoming launches of the China Space Station Telescope and the Large Synoptic Survey Telescope 
   will provide datasets with higher resolution and wider survey coverage. 
   Our future work will focus on improving the accuracy and reliability of our model through these datasets, 
   and promoting the application of the model to a wider range of astronomical applications. 
   We will also build on this foundation to conduct more in-depth research and provide solid 
   support for further exploration of astrophysics.

\begin{acknowledgements}
   We thank the anonymous referee for useful comments that helped us improve the 
   manuscript substantially. 
   This work is supported by the National Natural Science Foundation of China (NSFC) 
   under grant No. 11873037,
   the science research grants from the China Manned Space Project with No. CMS-CSST-2021-B05 
   and CMS-CSST-2021-A08, 
   the Natural Science Foundation of Shandong Province under grant No. ZR2022MA076,
   and partially supported by the Young Scholars Program of Shandong University, 
   Weihai (2016WHWLJH09) and GHfund A (202202018107).
   X.K is supported by the National Natural Science Foundation of China under grant No. 11803016.
   Z.Y is supported by the Shandong Province Natural Science Foundation grant No. ZR2022MA089.
\end{acknowledgements}

\appendix 

\section{proof}\label{appen A}

First, we analyze the original attention mechanism. In the original attention, the \textit{softmax} function is adopted hoping to achieve the 
"winner-takes-all" effect. However, there is an issue of attention dispersion in the interval $[0,1]$.
Consider given \(x_1, x_2 \in [0, 1]\) and $x_1 > x_2$. 
After being passed through the \textit{softmax}, the relative ratio of the two becomes \( e^{x_1} / e^{x_2} \). We define the attention efficiency, 
\(E_1\), as:
\begin{equation}
   E_1 = \frac{e^{x_1}}{e^{x_2}} / \frac{x_1}{x_2} = \frac{e^{x_1} / x_1}{e^{x_2} / x_2}
\end{equation}
It is easy to observe that \(e^x / x\) is monotonically decreasing in the interval [0, 1]. 
Since \(x_1 > x_2\), this implies \(E_1 < 1\). The \textit{softmax} function weakens the original weight ratio, causing the attention matrix to tend 
toward averaging. 
This becomes more pronounced when regularization strategies are employed.
In reality, the Taylor series expansion of \(e^x\) at the origin is:
\begin{equation}
   e^x = 1 + x + \frac{x^2}{2!} + \frac{x^3}{3!} + \dots + \frac{x^n}{n!} + R_n(x)
\end{equation}
It's evident that as \(x \to 0\), the ratio \(e^{x_1} / e^{x_2}\) is predominantly determined by the constant 1. 
Especially in the interval [0, 0.69], the constant 1 determines more than 50\% of the relative ratio. 
Now, we consider a new function, named \textit{enhanced\_softmax}.
To mitigate the interference caused by 1, we define:
\begin{equation}
   \textit{enhanced\_softmax} = \frac{e^{x_i}-1}{\sum_{j=1}^{n} (e^{x_j}-1)}
\end{equation}
where \(i = 1, 2,..., n\) is the \(i^{th}\) element of an attention vector of length \(n\).
Now, let's consider the attention efficiency of this operator, \(E_2\):
\begin{equation}
   E_2 = \frac{e^{x_1}-1}{e^{x_2}-1} / \frac{x_1}{x_2}
\end{equation}
At this point, \(E_2 = (e^x - 1) / x\) is monotonically increasing in the interval [0, 1] without altering the monotonicity in other regions 
relative to the \textit{softmax} function. 
Furthermore, when \(x_1 > x_2\), \(E_2 > 1\). This implies that the \textit{enhanced\_softmax} ensures that the self-attention mechanism retains
the "winner-takes-all" characteristic in the interval [0, 1].
Now, considering the attention efficiency of \textit{enhanced\_softmax} relative to \textit{softmax} over the interval $(0, \infty)$, denoted as \(E_3\), 
and taking \(x_3, x_4 \in (0, \infty)\) and $x_3 > x_4$:
\begin{equation}
   E_3 = \frac{e^{x_3}-1}{e^{x_4}-1} / \frac{e^{x_3}}{e^{x_4}} 
   = 1 + \frac{e^{x_3} - e^{x_4}}{e^{x_3 + x_4} - e^{x_3}} > 1
\end{equation}
Thus, across the entire range $(0, \infty)$, the \textit{enhanced\_softmax} always exhibits a stronger attention efficiency compared to the 
\textit{softmax} function, and it doesn't suffer from attention dispersion in the interval [0, 1].

\bibliographystyle{aa} 
\bibliography{sample631} 
\end{document}